\documentclass[%
 reprint,
 amsmath,amssymb,
 aps,
]{revtex4-2}

\usepackage{multirow}
\usepackage{booktabs}
\usepackage{graphicx}
\usepackage{dcolumn}
\usepackage{bm}
\usepackage{siunitx}
\usepackage[colorlinks=true, allcolors=blue]{hyperref} 
\allowdisplaybreaks

\begin{document}

\preprint{APS/123-QED}

\title{Stiffness matrix method for modelling wave propagation in arbitrary multilayers}

\author{Ming Huang, Frederic Cegla, Bo Lan}
 \email{Corresponding author: bo.lan@imperial.ac.uk}
\affiliation{Department of Mechanical Engineering, Imperial College London, London SW7 2AZ, United Kingdom}


\begin{abstract}

Natural and engineered media usually involve combinations of solid, fluid and porous layers, and accurate and stable modelling of wave propagation in such complex multilayered media is fundamental to evaluating their properties with wave-based methods. Here we present a general stiffness matrix method for modelling waves in arbitrary multilayers. The method first formulates stiffness matrices for individual layers based on the governing wave equations for fluids and solids, and the Biot theory for porous materials. Then it utilises the boundary conditions considered at layer interfaces to assemble the layer matrices into a global system of equations, to obtain solutions for reflection and transmission coefficients at any incidence. Its advantage over existing methods is manifested by its unconditional computational stability, and its validity is proved by experimental validations on single solid sheets, porous layers, and porous-solid-porous battery electrodes. This establishes a powerful theoretical platform that allows us to develop advanced wave-based methods to quantitatively characterise properties of the layers, especially for layers of porous materials.

\end{abstract}

\maketitle


\section{Introduction}\label{sec:intro}

Multilayered media are ubiquitous in nature as well as in engineering structures, with examples spanning from minerals and the Earth's crust to composite laminates and electrochemical systems (such as batteries). The layers normally consist of different material types, commonly involving solids and oftentimes fluid and porous layers. The resulting structures are generally complex, with a representative case being the electrodes of lithium-ion batteries, with two fluid-saturated porous layers coated on a thin solid metal sheet.

Consequently, multilayered media typically exhibit unique structural and functional properties. Achieving and maintaining the properties relies strongly on non-destructive methods to evaluate them and to monitor their changes. Ultrasonic testing is frequently used for this purpose and has facilitated many application areas, such as the estimation of the thicknesses of thin layered sheets \cite{Hagglund2009,Lerch2021}, the inspection of composite laminates \cite{Smith2018,Yang2021}, and the characterisation of the layered structures of lithium-ion batteries \cite{Copley2021,Huang2022b}. Such evaluations generally utilise the information about the layers that the ultrasonic waves carry after interacting with them. Therefore, for an ultrasonic method to deliver optimal results, understanding the wave interactions with the layered media
 through physical models is essential.

Matrix formulations are most commonly used for such models with arbitrary numbers of layers. As the earliest formulation, the transfer matrix method \cite{Thomson1950,Haskell1953} relates the stresses and displacements at one interface of a layer to those at the other interface. With continuity at the interfaces considered, the method produces a matrix for the entire system by multiplying the matrices of individual layers. Solving the final matrix equation in different ways can deliver solutions for wave reflections and transmissions as well as guided waves in the system. However, it suffers from computational instability at large $fd$ ($f$ is frequency and $d$ layer thickness) as inhomogeneous evanescent waves arise \cite{Potel1993,Castainqs1994}. To resolve this problem, a number of alternative formulations were proposed, and the global matrix method \cite{Knopoff1964,Randall1967} emerged as one of the preferred substitutes. Instead of using matrix multiplications, the global matrix method assembles the transfer matrices of all layers into a global system of equations. Its stability is achieved by eliminating the diverging exponential terms using different spatial origins for the partial waves \cite{Schmidt1985a,Schmidt1985b}. Another attractive approach to achieving computational stability at large $fd$ is the stiffness matrix method \cite{Rokhlin2002,Kausel1981}, which uses a stiffness matrix to link the stresses at the two interfaces of a layer to the respective displacements. This method can be implemented in a recursive form \cite{Rokhlin2002} similar to the matrix multiplication of the transfer matrix method, and can also be formulated in a global matrix form \cite{Kausel1981} in the same fashion as for transfer matrices. Both forms are unconditionally computationally stable. These matrix-based methods have received numerous applications in various fields, most notably in seismology \cite{Kennett2009}, ocean acoustics \cite{Jensen2011}, composites \cite{Nayfeh1995,rokhlin2011} and guided ultrasonics \cite{Lowe1995,Pavlakovic1997}.

Although the matrix formulations focused mainly on solid layers and occasionally on fluid ones (e.g. \cite{Kundu1985,Cervenka1991}), significant attention also centred on the development of matrix descriptions for porous layers. The formulations were mostly based on the Biot theory \cite{Biot1956a,Biot1956b,Biot1957,Biot1962} to describe the complex wave mechanics in fluid-saturated porous media. They all utilised transfer matrices to model the Biot waves in individual porous layers, but relied differently on matrix multiplication and global matrices to assemble layer matrices. The matrix multiplication method only applies to layered systems containing pure porous layers \cite{Allard1989,Jocker2004} or alternating fluid/solid-porous layers \cite{Lauriks1991}, while the global matrix method is a more general model for arbitrarily stacked fluid, solid and porous layers \cite{Brouard1995,Allard2009}. These developments have seen applications in, e.g., seismology \cite{Pride2002} and sound-absorbing materials \cite{Allard2009}. They are particularly useful for the inverse determination of important properties (porosity, tortuosity etc.) for porous materials \cite{Jocker2009,Allard2009}, and compared to non-matrix based inversion studies that are limited to single/double-layered settings \cite{Nagy1990,Fellah2007,Fellah2010}, they can deal with more complex cases with many layers of different types. However, the aforementioned instability problem arises, not only in the instability-prone matrix multiplication method but also to the supposedly-stable global matrix method. The problem in the former case occurs constantly at large $fd$ \cite{Jocker2004}, while that of the latter case, according to our analyses, arises less predictably at large incident angles of porous layers.

In this work, we present an intrinsically-stable stiffness matrix method for layered media with arbitrary numbers of fluid, solid and porous layers. The novelties and advantages are threefold. Firstly, the proposed method employs stiffness matrices to describe individual layers and uses global matrices to model assembled layers. Owing to the superior stability of both formulations, the proposed method exhibits intrinsic computational stability, and most importantly, it works exceptionally well for the cases that challenged existing methods. This allows us to reliably model highly-transmissible waves and guided modes that involve large wave angles in porous layers. Secondly, the proposed method is optimised to have simple expressions for both stiffness matrices and boundary conditions even for complex porous layers, thus enabling much easier computer implementation. Lastly, the proposed method is validated against experimental measurements to be working well for arbitrary single solid sheets, porous layers, and porous-solid-porous combinations. Based on the contributions, advanced ultrasonic techniques may be developed to characterise the properties of layered media, and an imperative application is to quantify the performance determinants of porous electrodes in lithium-ion batteries.

The paper is organised as follows. Section \ref{sec:theory} provides a concise review of the well-established wave physics in different layer materials. Then Sec. \ref{sec:method} presents the proposed stiffness matrix method, demonstrating how the wave physics in individual layers are modelled by stiffness matrices and how the layer matrices are assembled into global matrices to obtain wave solutions. This is followed by experimental validations in Sec. \ref{sec:results} to showcase the applicability of the method to complex layered media with a single solid/porous layer and multiple porous-solid-porous layers. Section \ref{sec:summary} concludes this paper.

\section{Wave physics in individual layers}\label{sec:theory}

\begin{figure}[b]
    \centering
    \includegraphics[width=0.8\linewidth]{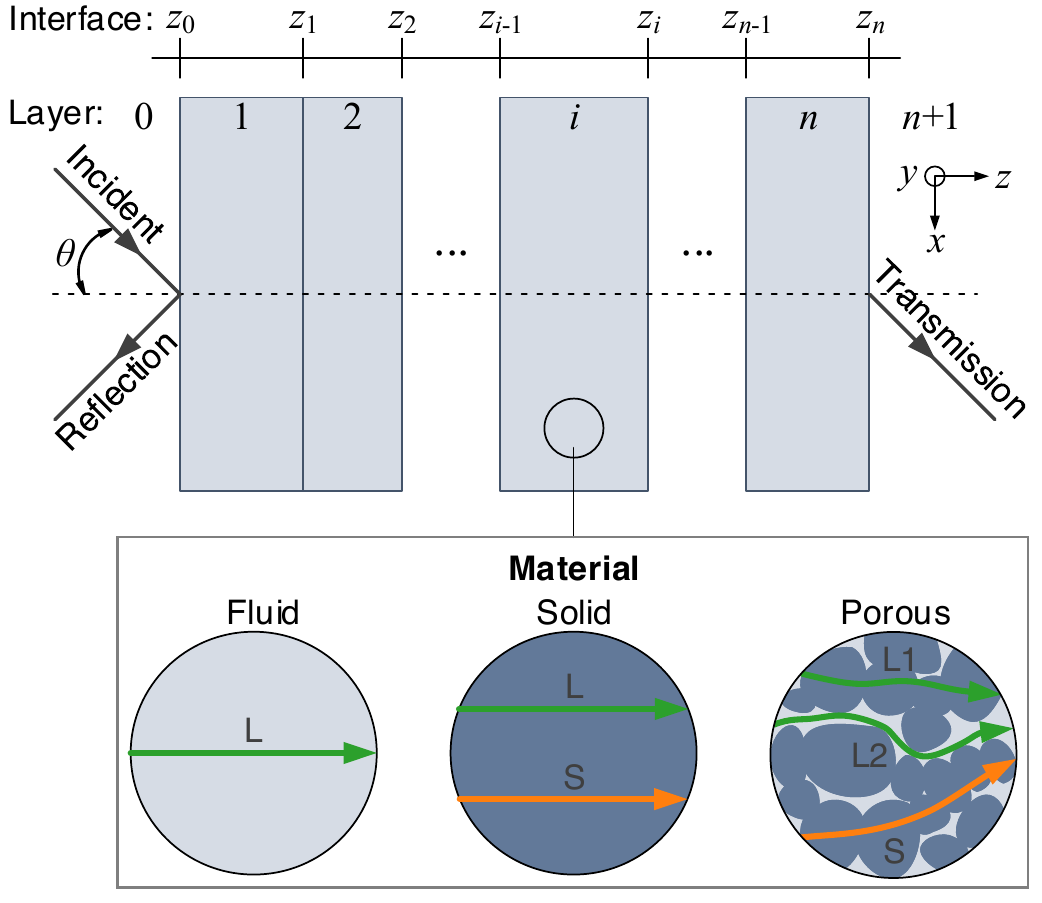}
    \caption{Wave propagation in a multilayered medium with fluid, solid and porous layers. The medium has $n$ layers and is bounded by half-spaces 0 and $n+1$. An incident wave from 0 induces a reflected wave back into 0 and a wave transmitted through the layers to $n+1$. Layer $i$ is defined by the interfaces of $z_{i-1}$ and $z_i$ and the thickness of $d_i=z_i-z_{i-1}$. Each layer is consisted of a fluid, solid or fluid-saturated porous material. There is one wave (longitudinal L) in fluid, two waves (longitudinal L and shear S) in solid, and three (fast L1 and slow L2 longitudinal, and shear) in a porous material.}
    \label{fig:illustrations}
\end{figure}

We address a general problem of wave propagation in an arbitrary multilayer, as illustrated in Fig. \ref{fig:illustrations}. The medium contains $n$ layers with infinite dimensions in the $x$- and $y$-directions, and layer $i$ is defined by interfaces $z_{i-1}$ and $z_i$ in the $z$-direction with thickness $d_i=z_i-z_{i-1}$. The layered system is bounded by half-spaces $0$ and $n+1$ on the two sides.

Individual layers in the system are each occupied by a fluid, elastic solid or fluid-saturated porous material. All three types of materials are treated as macroscopically isotropic and homogeneous. Wave propagation in individual layers is governed by different wave physics, depending on the nature of the layer material. The well-established governing equations in the three considered materials are reviewed in this section, and the stiffness matrix method will be formulated based upon them in the next.

\subsection{Fluid and solid layers}

We start with fluid and solid layers that involve relatively simple wave physics. With linear elasticity assumed and body forces neglected, wave propagation in fluid and solid materials is governed by the wave equation \cite{Graff1975}
\begin{equation}\label{eq:wave_equation}
    \nabla\cdot\bm{\sigma}-\rho(\partial^2\mathbf{u}/\partial t^2)=0,
\end{equation}
where $\mathbf{u}(\mathbf{x},t)$ and $\bm{\sigma}(\mathbf{x},t)$ are the particle displacement field and the stress tensor, both as function of the position $\mathbf{x}$ and time $t$. $\rho$ is 
the mass density of the material. $\nabla$ denotes the vector differential operator, namely $\nabla=[\partial /\partial x, \partial /\partial y, \partial /\partial z]^\mathrm{T}$. The stress tensor $\bm{\sigma}$ is related to the strain tensor $\bm{\varepsilon}$ by the generalised Hooke's law, given differently for fluid and solid materials by
\begin{eqnarray}
    p &=& K\varepsilon_{kk},\label{eq:stress_strain_fluid}\\
    \sigma_{ij} &=& (K-2G/3)\varepsilon_{kk}\delta_{ij}+2G\varepsilon_{ij},\label{eq:stress_strain_solid}
\end{eqnarray}
where fluid has an omnidirectional stress $p$ (also known as pressure) as a result of dilatation $\varepsilon_{kk}$, while solid exhibits direction-dependent stress $\sigma_{ij}$ ($i,j\in\{x,y,z\}$) due to dilatation $\varepsilon_{kk}$ and shearing $\varepsilon_{ij}$ ($i\neq j$) in the medium. Note that Einstein summation over the repeated index $k$ from $x$ to $z$ is assumed for the dilatation $\varepsilon_{kk}$. $\delta_{ij}$ is the Kronecker delta. $K$ and $G$ are the bulk and shear moduli. The strain component $\varepsilon_{ij}$ and dilatation $\varepsilon_{kk}$ are related to the displacement field by
\begin{eqnarray}
    \varepsilon_{ij}&=&(\partial u_i/\partial x_j+\partial u_j/\partial x_i)/2,\label{eq:strain} \\
    \varepsilon_{kk}&=&\varepsilon_{xx}+\varepsilon_{yy}+\varepsilon_{zz}=\nabla\cdot\mathbf{u}.\label{eq:dilatation}
\end{eqnarray}

Substituting the above strain-displacement relation into the Hooke's law and then into Eq. \ref{eq:wave_equation} leads to an equation for the displacement field $\mathbf{u}$. The equation can be further written for the scalar $\varphi$ and vector $\mathbf{H}$ potentials by using the Helmholtz decomposition \cite{Morse1953,Graff1975}
\begin{equation}\label{eq:helmholtz}
    \mathbf{u}=\nabla\varphi+\nabla\times\mathbf{H}.
\end{equation}
The two potentials describe respectively the longitudinal (dilatational) and shear (rotational) waves in the medium. For fluids, the vector potential $\mathbf{H}$ vanishes due to the absence of shear waves, and solving the resulting wave equation for the scalar potential delivers a longitudinal wave solution with wave speed 
\begin{equation}
    c_\mathrm{L}=\sqrt{K/\rho}.
\end{equation}
For solids, the wave equation is decoupled into two equations for the scalar and vector potentials respectively. The two equations give respectively the longitudinal and shear wave solutions, having the wave speeds of
\begin{equation}
    c_\mathrm{L}=\sqrt{(K+4G/3)/\rho}, \: c_\mathrm{S}=\sqrt{G/\rho}.
\end{equation}
Note we have conveniently treated the two differently-polarised shear waves as a single wave mode because they have the the same speed in the considered isotropic solid. This applies to the porous media as discussed below.

\subsection{Fluid-saturated porous layers}

Now we consider fluid-saturated porous layers. The wave physics are much more complicated in this case and are addressed by the widely-employed Biot theory \cite{Biot1956a,Biot1956b,Biot1957} (or empirically by other models such as \cite{Horoshenkov2016,Horoshenkov2019}). Here, the solid frame is considered to be continuous, and the pores fully connected and saturated with fluid. The propagating wave is subjected to attenuation induced by scattering in the solid phase, and viscous and inertial dissipation in the fluid phase. When the wavelength is large compared to the average pore size, the propagating wave can be treated in a homogenised sense. The average displacement fields can then be characterised by $\mathbf{u}^\mathrm{s}(\mathbf{x},t)$ and $\mathbf{u}^\mathrm{f}(\mathbf{x},t)$ in the solid (`s') and fluid (`f') phases. The two wave fields are coupled and described by the wave equations \cite{Biot1956a,Johnson1982a}
\begin{eqnarray}
    \nabla\cdot\bm{\sigma}^\mathrm{s}-\frac{\partial^2}{\partial t^2}(\rho_{11}\mathbf{u}^\mathrm{s}+\rho_{12}\mathbf{u}^\mathrm{f})\label{eq:wave_equation_porous1}\\ \nonumber
    -bF\frac{\partial}{\partial t}(\bm{u}^\mathrm{s}-\bm{u}^\mathrm{f})=0,\\
    \nabla\cdot\bm{\sigma}^\mathrm{f}-\frac{\partial^2}{\partial t^2}(\rho_{12}\mathbf{u}^\mathrm{s}+\rho_{22}\mathbf{u}^\mathrm{f})\label{eq:wave_equation_porous2}\\ \nonumber
    -bF\frac{\partial}{\partial t}(\bm{u}^\mathrm{f}-\bm{u}^\mathrm{s})=0,
\end{eqnarray}
where $\rho_{11}$, $\rho_{12}$ and $\rho_{22}$ are the effective densities \cite{Biot1956a}
\begin{eqnarray}
    \rho_{12}&=&-(\alpha_\infty-1)\phi\rho_\mathrm{f}, \\
    \rho_{11}&=&(1-\phi)\rho_\mathrm{s}-\rho_{12}, \\
    \rho_{22}&=&\phi\rho_\mathrm{f}-\rho_{12},
\end{eqnarray}
where $\rho_\mathrm{s}$ and $\rho_\mathrm{f}$ are the densities of the solid and fluid materials. $\phi$ is the porosity and $\alpha_\infty$ the tortuosity. The parameter $b$ in the wave equations represents the viscous damping factor, given by \cite{Johnson1982a,Jocker2009}
\begin{equation}
    b=\eta \phi^2/k_0
\end{equation}
with $\eta$ being the dynamic viscosity of the fluid and $k_0$ the permeability of the fluid through the porous medium. $F$ is the viscous correction factor with a generalised form of \cite{Biot1956a,Johnson1982a,Johnson1987}
\begin{equation}
    F=\sqrt{1+iMf/(2f_c)},
\end{equation}
which is dependent on the frequency $f$. $M$ is the shape factor, which is generally taken as unity. $f_c$ is the viscous characteristic frequency, given by \cite{Johnson1987,Jocker2009}
\begin{equation}
    f_c=\eta\phi/(2\pi\alpha_\infty\rho_\mathrm{f}k_0).
\end{equation}

In Eqs. \ref{eq:wave_equation_porous1} and \ref{eq:wave_equation_porous2}, the stress tensors $\bm{\sigma}^\mathrm{s}$ and $\bm{\sigma}^\mathrm{f}$ in the solid and fluid phases are related to the strain tensors by \cite{Biot1956a,Johnson1982a}
\begin{eqnarray}
    \sigma_{ij}^\mathrm{s}&=&[(P-2N)\varepsilon_{kk}^\mathrm{s}+Q\varepsilon_{kk}^\mathrm{f}]\delta_{ij}+2N\varepsilon_{ij}^\mathrm{s},\label{eq:stress_strain_porous1}\\
    \sigma_{ij}^\mathrm{f}&=&[Q\varepsilon_{kk}^\mathrm{s}+R\varepsilon_{kk}^\mathrm{f}]\delta_{ij},\label{eq:stress_strain_porous2}
\end{eqnarray}
with the strain tensors linked to the respective displacement fields in the solid and fluid by Eqs. \ref{eq:strain} and \ref{eq:dilatation}. $P$ and $N$ are the effective longitudinal and shear moduli of the medium. $R$ represents the pressure required for forcing a certain volume of the liquid into the medium whilst maintaining the total volume. $Q$ signifies the coupling of volume change between the solid and liquid. These four elastic parameters are given by \cite{Biot1957,Jocker2009}
\begin{eqnarray}
        P&=&K_\mathrm{b}+K_\mathrm{f}(1-\phi-K_\mathrm{b}/K_\mathrm{s})^2/\phi_\mathrm{eff}+4G_\mathrm{b}/3, \\
        N&=&G_\mathrm{b}, \\
        R&=&\phi^2K_\mathrm{f}/\phi_\mathrm{eff}, \\
        Q&=&\phi K_\mathrm{f}(1-\phi-K_\mathrm{b}/K_\mathrm{s})/\phi_\mathrm{eff},
\end{eqnarray}
where $\phi_\mathrm{eff}=\phi+K_\mathrm{f}/K_\mathrm{s}(1-\phi-K_\mathrm{b}/K_\mathrm{s})$ is an effective porosity of the fluid-saturated medium. $K_\mathrm{s}$ and $K_\mathrm{f}$ are the bulk moduli of the solid and fluid materials, respectively. $K_\mathrm{b}$ and $G_\mathrm{b}$ are the in-vacuo bulk and shear moduli of the solid frame (namely, the porous solid after draining out the saturated fluid).

Using the Helmholtz decomposition, the wave equations in Eqs. \ref{eq:wave_equation_porous1} and \ref{eq:wave_equation_porous2} can be decoupled into two equations for longitudinal and shear waves, respectively \cite{Biot1956a}. The longitudinal wave equation delivers two solutions with the wave speeds of
\begin{eqnarray}
    c_\mathrm{L1}^2 &=& \frac{2(PR-Q^2)}{P\tilde{\rho}_{22}+R\tilde{\rho}_{11}-2Q\tilde{\rho}_{12}-\sqrt{\Delta}}, \\
    c_\mathrm{L2}^2 &=& \frac{2(PR-Q^2)}{P\tilde{\rho}_{22}+R\tilde{\rho}_{11}-2Q\tilde{\rho}_{12}+\sqrt{\Delta}},
\end{eqnarray}
where
\begin{eqnarray}
    \tilde{\rho}_{12} &=& \rho_{12}+ibF/\omega, \\
    \tilde{\rho}_{11} &=& \rho_{11}-ibF/\omega, \\
    \tilde{\rho}_{22} &=& \rho_{22}-ibF/\omega,\\
    \Delta &=& (P\tilde{\rho}_{22}+R\tilde{\rho}_{11}-2Q\tilde{\rho}_{12})^2 \\ \nonumber
    &&-4(PR-Q^2)(\tilde{\rho}_{11}\tilde{\rho}_{22}-\tilde{\rho}_{12}^2).
\end{eqnarray}
The two longitudinal waves both involve coupled motion in the solid frame and the saturated fluid. The faster wave L1 propagates dominantly in the solid frame, with a speed slower than that of the solid and faster than the saturated fluid. The slower wave L2 travels predominantly in the fluid phase, having a speed slower than the fluid. The shear wave equation yields only one solution involving coupled motion between the solid and fluid, with the wave speed of
\begin{equation}
    c_\mathrm{S}^2=N\tilde{\rho}_{22}/(\tilde{\rho}_{11}\tilde{\rho}_{22}-\tilde{\rho}_{12}^2).
\end{equation}

\section{Stiffness matrix method}\label{sec:method}

With the wave physics in individual layers discussed, here we address the propagation of waves in the entire layered system by formulating the stiffness matrix method.

\subsection{Stiffness matrix for the two interfaces of a layer}

The formulation begins by establishing a stiffness matrix relation for a layer $i$ by \cite{Rokhlin2002}
\begin{equation}\label{eq:relation_stiffness}
\begin{bmatrix} \bm{\sigma}_{i-1} \\ \bm{\sigma}_{i} \end{bmatrix}^i = \mathbf{K}^i \begin{bmatrix} \mathbf{u}_{i-1} \\ \mathbf{u}_{i} \end{bmatrix}^i,
\end{equation}
which relates the stress vectors $\bm{\sigma}$ on the two interfaces to the respective displacement vectors $\mathbf{u}$ by the stiffness matrix $\mathbf{K}$. Note that layer $i$ is bounded by the interfaces $i-1$ and $i$ (see Fig. \ref{fig:illustrations}), and layers and interfaces are differently indicated by superscripts and subscripts throughout this paper wherever possible. The displacement and stress vectors each has $m$ components that are representative of the $m$ wave modes in the layer. As clarified in the preceding section, the fluid, solid and porous layers considered in this work have $m=1$, 2 and 3 wave modes, respectively. Here we emphasise again that the two differently-polarised shear waves are treated as a single shear wave mode for solids and porous materials. We choose the displacement and stress vectors as
\begin{align}
    &\mathrm{Fluid:}  && \mathbf{u}=\left[ u_z \right]^\mathrm{T}, \: &&\bm{\sigma}=\left[ p \right]^\mathrm{T}, \label{eq:vectors_fluid} \\
    &\mathrm{Solid:}  && \mathbf{u}=\left[ u_z, u_x \right]^\mathrm{T}, \: &&\bm{\sigma}=\left[ \sigma_{zz}, \sigma_{xz} \right]^\mathrm{T}, \label{eq:vectors_solid} \\
    &\mathrm{Porous:} && \mathbf{u}=\left[ u_z, u_x^\mathrm{s}, \hat{u}_z \right]^\mathrm{T}, \: &&\bm{\sigma}=\left[ p, \sigma_{xz}^\mathrm{s}, \hat{\sigma}_{zz} \right]^\mathrm{T}, \label{eq:vectors_porous}
\end{align}
where the dependencies on space $\{x,y,z\}$ and time $t$ are implied. For porous layers, the components $u_x^\mathrm{s}$ and $\sigma_{xz}^\mathrm{s}$ are for the solid frame; the other four components, however, contain the displacements and stresses of both the solid and fluid phases, given by
\begin{align}
    u_z               &=(1-\phi)u_z^\mathrm{s} + \phi u_z^\mathrm{f}, \label{eq:vectors_porous1}\\
    \hat{u}_z         &=u_z^\mathrm{s} - u_z^\mathrm{f}, \label{eq:vectors_porous2}\\
    p                 &=\sigma_{zz}^\mathrm{s}+\sigma_{zz}^\mathrm{f}, \label{eq:vectors_porous3}\\
    \hat{\sigma}_{zz} &=\sigma_{zz}^\mathrm{s}/(1-\phi)-\sigma_{zz}^\mathrm{f}/\phi. \label{eq:vectors_porous4}
\end{align}
which are chosen to ease the definition of boundary conditions in the next subsection; this will be discussed in detail below.

To obtain the displacement and stress vectors, we write the scalar $\varphi$ and vector $\mathbf{H}$ potentials in a layer as \cite{Lowe1995,Jocker2004}
\begin{align}
    &\mathrm{Fluid:}  && \varphi=U_\mathrm{L}, \: \mathbf{H}=\mathbf{0}, \\
    &\mathrm{Solid:}  && \varphi=U_\mathrm{L}, \: \mathbf{H}=\left[ 0, U_\mathrm{S}, 0 \right]^\mathrm{T}, \\
    &\mathrm{Porous:} && \varphi^\mathrm{s}=U_\mathrm{L1}+U_\mathrm{L2}, \: \mathbf{H}^\mathrm{s}=\left[ 0, U_\mathrm{S}, 0 \right]^\mathrm{T}, \\
    &                 && \varphi^\mathrm{f}=\mu_\mathrm{L1}U_\mathrm{L1}+\mu_\mathrm{L2}U_\mathrm{L2}, \: \mathbf{H}^\mathrm{s}=\mu_\mathrm{S}\mathbf{H}^\mathrm{s},
\end{align}
with
\begin{equation}\label{eq:wave_component}
    U_{i} = (a_{i}^+e^{ik_{iz}z}+a_{i}^-e^{-ik_{iz}z})e^{i(k_{x}x-\omega t)},
\end{equation}
where $i\in\{\mathrm{L},\mathrm{L1},\mathrm{L2},\mathrm{S}\}$. Here each scalar potential corresponds to a longitudinal wave mode and each vector potential to a shear mode. Without loss of generality, the nonzero components in the vector potentials are assumed to be in the $y$ direction, meaning that the particle motion of the shear waves lies in the $x$-$z$ plane. Each wave mode travels in both the forward (`+') and backward (`--') directions, and these two waves are represented by the two terms in Eq. \ref{eq:wave_component} with amplitudes $a^+$ and $a^-$. In addition, it is implied in the potentials that the incident wave has a wavenumber component of $k_x$ in the transverse $x$-direction of the layered system. According to the Snell's law, all propagating waves in all layers have the same transverse component of wavenumber. As a result, the common term $e^{i(k_{x}x-\omega t)}$ in Eq. \ref{eq:wave_component} is invariant, and the wavenumber component in the $z$-direction for a wave $i$ ($i\in\{\mathrm{L},\mathrm{L1},\mathrm{L2},\mathrm{S}\}$) is given by
\begin{equation}
    k_{iz}=\sqrt{k_{i}^2-k_{x}^2},
\end{equation}
where the total wavenumber $k_i$ is related to the wave speed $c_i$ by $k_i=\omega/c_i$, with $\omega=2\pi f$ being the angular frequency. Also, we have to emphasise that the potentials are given separately for the solid and fluid phases in a porous medium, in the same fashion as for the displacement and stress fields in Sec. \ref{sec:theory}. The potentials for the two phases are related by the ratios $\mu_i$ due to the coupling of the wave fields between the two phases. The ratios can be obtained from the physical relations in Sec. \ref{sec:theory}, given by \cite{Jocker2009}
\begin{eqnarray}
    \mu_\mathrm{L1} &=& (\tilde{\rho}_{11}-P/c_\mathrm{L1}^2)/(Q/c_\mathrm{L1}^2-\tilde{\rho}_{12}),\\
    \mu_\mathrm{L2} &=& (\tilde{\rho}_{11}-P/c_\mathrm{L2}^2)/(Q/c_\mathrm{L2}^2-\tilde{\rho}_{12}),\\
    \mu_\mathrm{S} &=& -\tilde{\rho}_{12}/\tilde{\rho}_{22}.
\end{eqnarray}

In the potentials, the $z$-coordinate origin of each wave is defined to be at its entry to the layer. Therefore, forward propagating waves (L+, L1+, L2+, S+) in a layer $i$ have their origin at $z_{i-1}$ and backward propagating waves (L--, L1--, L2--, S--) have their origin at $z_{i}$. Such selection ensures that every single exponential term in the potentials is normalised to unity on its entry interface and decays towards the exit interface. This essentially eliminates the numerical overflow of the exponential terms as the waves become inhomogeneous, thus impeding numerical instability \cite{Lowe1995,Rokhlin2002}.

Based on the potentials, the components of the displacement vector $\mathbf{u}$ in Eqs. \ref{eq:vectors_fluid}-\ref{eq:vectors_porous} are obtained using the Helmholtz decomposition (Eq. \ref{eq:helmholtz}).  For a given layer $i$, the displacement vector can then be evaluated at its two interfaces $z_{i-1}$ and $z_{i}$, leading to a matrix-form result
\begin{equation}\label{eq:relation_displacement}
    \begin{bmatrix} \mathbf{u}_{i-1} \\ \mathbf{u}_{i} \end{bmatrix}^i = 
    \begin{bmatrix} \mathbf{D}^+ & \mathbf{D}^-\mathbf{E} \\ \mathbf{D}^+\mathbf{E} & \mathbf{D}^- \end{bmatrix}^i
    \begin{bmatrix} \mathbf{A}^+ \\ \mathbf{A}^- \end{bmatrix}^i = \mathbf{D}^i\mathbf{A}^i,
\end{equation}
where a common term $ie^{i(k_{x}x-\omega t)}$ is implied. $\mathbf{D}$ is the displacement matrix, and the expressions of its $m\times m$ sub-matrices $\mathbf{D}^+$ and $\mathbf{D}^-$ are provided in Table \ref{tab:matrices} for the three material types. $\mathbf{E}$ is a $m\times m$ diagonal matrix, with the diagonal being $\left[e^{ik_{\mathrm{L}z}d}\right]^\mathrm{T}$, $\left[e^{ik_{\mathrm{L}z}d},e^{ik_{\mathrm{S}z}d}\right]^\mathrm{T}$ and $\left[e^{ik_{\mathrm{L1}z}d},e^{ik_{\mathrm{L2}z}d},e^{ik_{\mathrm{S}z}d}\right]^\mathrm{T}$ for fluid, solid and porous materials. $\mathbf{A}$ is the amplitude vector, and its sub-vectors are $\mathbf{A}^\pm=\left[a_\mathrm{L}^\pm\right]^\mathrm{T}$, $\left[a_\mathrm{L}^\pm,a_\mathrm{S}^\pm\right]^\mathrm{T}$ and $\left[a_\mathrm{L1}^\pm,a_\mathrm{L2}^\pm,a_\mathrm{S}^\pm\right]^\mathrm{T}$ for the three types of materials.

\begin{table}[b]
\footnotesize
\caption{Sub-matrices of the displacement $\mathbf{D}$ and stress $\mathbf{S}$ matrices for different types of layer materials. For the porous case, $h_i=\mu_i-1$, $g_i=1+\phi h_i$, $q_i=k_i^2(P+\mu_iQ)/N-2k_x^2$, $r_i=k_i^2(Q+\mu_iR)/N$, $s_i=[\phi q_i+(\phi-1)r_i]/[\phi(\phi-1)]$ for $i\in\{\mathrm{L1},\mathrm{L2},\mathrm{S}\}$.}
\label{tab:matrices}
\begin{ruledtabular}
\begin{tabular}{ll}
\textrm{Layer} & \textrm{Matrix}\\ \colrule \addlinespace[1ex]
\multirow{2}{*}{Fluid}  & $\mathbf{D}^+=\begin{bmatrix} k_{\mathrm{L}z} \end{bmatrix}$, \:
                          $\mathbf{D}^-=\begin{bmatrix} -k_{\mathrm{L}z} \end{bmatrix}$ \\ \addlinespace[1ex]
                        & $\mathbf{S}^+=\begin{bmatrix} i\rho\omega^2 \end{bmatrix}$, \:
                          $\mathbf{S}^-=\begin{bmatrix} i\rho\omega^2 \end{bmatrix}$ \\ \colrule \addlinespace[1ex]
\multirow{6}{*}{Solid}  & $\mathbf{D}^+=\begin{bmatrix} k_{\mathrm{L}z} & k_x \\ k_x & -k_{\mathrm{S}z} \end{bmatrix}$, \:
                         $\mathbf{D}^-=\begin{bmatrix} -k_{\mathrm{L}z} & k_x \\ k_x & k_{\mathrm{S}z} \end{bmatrix}$ \\ \addlinespace[1ex]
                        & $\mathbf{S}^+=i\rho c_\mathrm{S}^2\begin{bmatrix} k_\mathrm{S}^2-2k_x^2 &  2k_xk_{\mathrm{S}z} \\  2k_xk_{\mathrm{L}z} & -(k_\mathrm{S}^2-2k_x^2) \end{bmatrix}$ \\ \addlinespace[1ex]
                        & $\mathbf{S}^-=i\rho c_\mathrm{S}^2\begin{bmatrix} k_\mathrm{S}^2-2k_x^2 & -2k_xk_{\mathrm{S}z} \\ -2k_xk_{\mathrm{L}z} & -(k_\mathrm{S}^2-2k_x^2) \end{bmatrix}$ \\ \colrule \addlinespace[1ex]
\multirow{12}{*}{Porous} & $\mathbf{D}^+=\begin{bmatrix} g_\mathrm{L1}k_{\mathrm{L1}z} & g_\mathrm{L2}k_{\mathrm{L2}z} & g_\mathrm{S}k_x \\ k_x & k_x & -k_{\mathrm{S}z} \\ -h_\mathrm{L1}k_{\mathrm{L1}z} & -h_\mathrm{L2}k_{\mathrm{L2}z} & -h_\mathrm{S}k_x \end{bmatrix}$ \\ \addlinespace[1ex]
                        & $\mathbf{D}^-=\begin{bmatrix} -g_\mathrm{L1}k_{\mathrm{L1}z} & -g_\mathrm{L2}k_{\mathrm{L2}z} & g_\mathrm{S}k_x \\ k_x & k_x & k_{\mathrm{S}z} \\ h_\mathrm{L1}k_{\mathrm{L1}z} & h_\mathrm{L2}k_{\mathrm{L2}z} & -h_\mathrm{S}k_x \end{bmatrix}$ \\ \addlinespace[1ex]
                        & $\mathbf{S}^+=iN\begin{bmatrix} q_\mathrm{L1}+r_\mathrm{L1} &  q_\mathrm{L2}+r_\mathrm{L2} & 2k_xk_{\mathrm{S}z} \\ 2k_xk_{\mathrm{L1}z} & 2k_xk_{\mathrm{L2}z} & -(k_\mathrm{S}^2-2k_x^2) \\ -s_\mathrm{L1} & -s_\mathrm{L2} & 2k_xk_{\mathrm{S}z}/(1-\phi) \end{bmatrix}$ \\ \addlinespace[1ex]
                        & $\mathbf{S}^-=iN\begin{bmatrix} q_\mathrm{L1}+r_\mathrm{L1} &  q_\mathrm{L2}+r_\mathrm{L2} & -2k_xk_{\mathrm{S}z} \\ -2k_xk_{\mathrm{L1}z} & -2k_xk_{\mathrm{L2}z} & -(k_\mathrm{S}^2-2k_x^2) \\ -s_\mathrm{L1} & -s_\mathrm{L2} & 2k_xk_{\mathrm{S}z}/(\phi-1) \end{bmatrix}$
\end{tabular}
\end{ruledtabular}
\end{table}

The components of the stress vector $\bm{\sigma}$ in Eqs. \ref{eq:vectors_fluid}-\ref{eq:vectors_porous} are obtained using the respective strain-displacement relations and the Hooke's laws in Sec. \ref{sec:theory}, yielding
\begin{equation}\label{eq:relation_stress}
    \begin{bmatrix} \bm{\sigma}_{i-1} \\ \bm{\sigma}_{i} \end{bmatrix}^i = 
    \begin{bmatrix} \mathbf{S}^+ & \mathbf{S}^-\mathbf{E} \\ \mathbf{S}^+\mathbf{E} & \mathbf{S}^- \end{bmatrix}^i
    \begin{bmatrix} \mathbf{A}^+ \\ \mathbf{A}^- \end{bmatrix}^i = \mathbf{S}^i\mathbf{A}^i,
\end{equation}
with $ie^{i(k_{x}x-\omega t)}$ implied. $\mathbf{S}$ is the stress matrix and its $m\times m$ sub-matrices $\mathbf{S}^+$ and $\mathbf{S}^-$ are listed in Table \ref{tab:matrices} for the three material types. The diagonal matrix $\mathbf{E}$ and the amplitude vector $\mathbf{A}$ are the same as those for the displacement vector in Eq. \ref{eq:relation_displacement}.

Equations \ref{eq:relation_displacement} and \ref{eq:relation_stress} relate the displacement and stress vectors on the layer interfaces to the amplitude vector. By obtaining the amplitude vector $\mathbf{A}^i$ from Eq. \ref{eq:relation_displacement} and substituting into Eq. \ref{eq:relation_stress}, we arrive at the stiffness matrix relation in Eq. \ref{eq:relation_stiffness}, with the stiffness matrix given by
\begin{align}\label{eq:matrix_stiffness}
    \mathbf{K}^i&=\begin{bmatrix} \mathbf{K}_{11}^i & \mathbf{K}_{12}^i \\ \mathbf{K}_{21}^i & \mathbf{K}_{22}^i \end{bmatrix} = \mathbf{S}^i\left(\mathbf{D}^i\right)^{-1}& \\
    &=\begin{bmatrix} \mathbf{S}^+ & \mathbf{S}^-\mathbf{E} \\ \mathbf{S}^+\mathbf{E} & \mathbf{S}^- \end{bmatrix}^i \left(
     \begin{bmatrix} \mathbf{D}^+ & \mathbf{D}^-\mathbf{E} \\ \mathbf{D}^+\mathbf{E} & \mathbf{D}^- \end{bmatrix}^i\right)^{-1}, \nonumber
\end{align}
where $\mathbf{K}_{pq}^i\,(p,q\in\{1,2\})$ are $m\times m$ sub-matrices.

\subsection{Boundary conditions across the interface of two layers}

After establishing the stiffness relation for the two interfaces of a layer, now we consider the wave interaction across the interface of two neighbouring layers. The interaction is defined by the boundary conditions, which can be expressed in matrix form for the displacement and stress vectors as
\begin{align}
    \mathbf{B}_i^{i}\mathbf{u}_i^{i}+\mathbf{B}_i^{i+1}\mathbf{u}_i^{i+1}=\mathbf{0}, \label{eq:bc_displacement}\\
    \mathbf{C}_i^{i}\bm{\sigma}_i^{i}+\mathbf{C}_i^{i+1}\bm{\sigma}_i^{i+1}=\mathbf{0}, \label{eq:bc_stress}
\end{align}
across the interface $i$ (subscript) between layers $i$ and $i+1$ (superscript); see Fig. \ref{fig:illustrations} for the numbering of layers and interfaces. $\mathbf{B}$ and $\mathbf{C}$ are the boundary matrices for the displacement and stress vectors.

The boundary matrices vary depending on the material types of the two neighbouring layers. When the two layers are made of the same material, the displacements and stresses of the two layers need to be continuous across the interface, resulting in the boundary matrices
\begin{equation}
    \mathbf{B}_i^{i}=-\mathbf{B}_i^{i+1}=\mathbf{C}_i^{i}=-\mathbf{C}_i^{i+1}=\mathbf{I}_{m\times m},
\end{equation}
where $\mathbf{I}_{m\times m}$ represents the identity matrix of size $m$.

When the two layers have different material types, they involve different numbers of wave modes $m$, leading to different numbers of components in the displacement and stress vectors. In this case, the boundary matrices are complicated by the fact that there are not only continuity conditions but also Dirichlet conditions for the displacement and/or stress components. Across a fluid-solid interface for example, beside the two continuity conditions $u_z^\mathrm{f}=u_z^\mathrm{s}$ and $p^\mathrm{f}=\sigma_{zz}^\mathrm{s}$, an extra Dirichlet condition exists for the shear stress of the solid layer (namely $\sigma_{xz}^\mathrm{s}=0$) due to the lack of shear stresses in the neighbouring fluid. These three conditions translate to the boundary matrices in Table \ref{tab:interfaces}, alongside those for other interface types.

Before proceeding, we shall revisit the choice of the displacement and stress components for porous layers in Eqs. \ref{eq:vectors_porous1}-\ref{eq:vectors_porous4}. As aforementioned, the choice makes it straightforward to define boundary conditions, specifically: (1) Equating $u_z$ in Eq. \ref{eq:vectors_porous1} to the $u_z$ of a neighbouring layer prescribes the conservation of fluid and solid volume through a fluid- or solid-porous interface. (2) Applying the Dirichlet condition to $\hat{u}_z$ in Eq. \ref{eq:vectors_porous2} ensures no mass is lost over a solid-porous interface. (3) The equality of $p$ in Eq. \ref{eq:vectors_porous3} to the $p$ or $\sigma_{zz}$ of a neighbouring fluid or solid layer satisfies the continuity of normal stress across the interface. (4) The use of the Dirichlet condition on $\hat{\sigma}_{zz}$ in Eq. \ref{eq:vectors_porous4} additionally guarantees the continuity of fluid pressure across a fluid-porous interface. These aspects deliver very simple boundary matrices as given in Table \ref{tab:interfaces} despite the complex wave physics in porous layers.

\begin{table}[b]
\footnotesize
\caption{Boundary matrices $\mathbf{B}$ and $\mathbf{C}$ for the displacement and stress vectors across the interface of two layers. The superscripts f, s and p indicate respectively fluid, solid and porous layers.}
\label{tab:interfaces}
\begin{ruledtabular}
\begin{tabular}{ll}
\textrm{Interface} & \textrm{Matrix}\\ \colrule \addlinespace[1ex]
\multirow{2}{*}{Fluid-solid}  & $\mathbf{B}^\mathrm{f}=\mathbf{I}_{1\times1}$, \,
                                $\mathbf{B}^\mathrm{s}=-\begin{bmatrix} 1 & 0 \end{bmatrix}$, \\ \addlinespace[1ex]
                              & $\mathbf{C}^\mathrm{f}=\begin{bmatrix} 1 & 0 \end{bmatrix}^\mathrm{T}$, \,
                                $\mathbf{C}^\mathrm{s}=-\mathbf{I}_{2\times2}$ \\ \colrule \addlinespace[1ex]
\multirow{2}{*}{Fluid-porous} & $\mathbf{B}^\mathrm{f}=\mathbf{I}_{1\times1}$, \,
                                $\mathbf{B}^\mathrm{p}=-\begin{bmatrix} 1 & 0 & 0 \end{bmatrix}$, \\ \addlinespace[1ex]
                              & $\mathbf{C}^\mathrm{f}=\begin{bmatrix} 1 & 0 & 0 \end{bmatrix}^\mathrm{T}$, \,
                                $\mathbf{C}^\mathrm{p}=-\mathbf{I}_{3\times3}$ \\ \colrule \addlinespace[1ex]
\multirow{4}{*}{Solid-porous}& $\mathbf{B}^\mathrm{s}=\begin{bmatrix} 1 & 0 & 0 \\ 0 & 1 & 0 \end{bmatrix}^\mathrm{T}$, \,
                                $\mathbf{B}^\mathrm{p}=-\mathbf{I}_{3\times3}$, \\ \addlinespace[1ex]
                              & $\mathbf{C}^\mathrm{s}=\mathbf{I}_{2\times2}$, \,
                                $\mathbf{C}^\mathrm{p}=-\begin{bmatrix} 1 & 0 & 0 \\ 0 & 1 & 0 \end{bmatrix}$
\end{tabular}
\end{ruledtabular}
\end{table}

\subsection{Global matrix for the calculation of reflection and transmission coefficients}

Substituting the stiffness matrix relation in Eq. \ref{eq:relation_stiffness} into the stress boundary condition in Eq. \ref{eq:bc_stress}, we have
\begin{equation}\footnotesize
    \mathbf{C}_i^{i}(\mathbf{K}_{21}^{i}\mathbf{u}_{i-1}^{i}+\mathbf{K}_{22}^{i}\mathbf{u}_{i}^{i})+\mathbf{C}_i^{i+1}(\mathbf{K}_{11}^{i+1}\mathbf{u}_{i}^{i+1}+\mathbf{K}_{12}^{i+1}\mathbf{u}_{i+1}^{i+1})=\mathbf{0},
\end{equation}
which gives the following result when combined with the displacement boundary condition in Eq. \ref{eq:bc_displacement} as
\begin{equation}\footnotesize
    \begin{bmatrix}
        \mathbf{0} & \mathbf{B}_i^{i} & \mathbf{B}_i^{i+1} & \mathbf{0} \\
        \mathbf{C}_i^{i}\mathbf{K}_{21}^{i} & \mathbf{C}_i^{i}\mathbf{K}_{22}^{i} & \mathbf{C}_i^{i+1}\mathbf{K}_{11}^{i+1} & \mathbf{C}_i^{i+1}\mathbf{K}_{12}^{i+1}
    \end{bmatrix}
    \begin{bmatrix}
        \mathbf{u}_{i-1}^{i} \\ \mathbf{u}_{i}^{i} \\ \mathbf{u}_{i}^{i+1} \\ \mathbf{u}_{i+1}^{i+1}
    \end{bmatrix} = \mathbf{0}.
\end{equation}
With this procedure applied to all interfaces of a layered system, a global matrix for the interfacial displacements can be obtained. We note that a recursive algorithm can be used instead if all layers belong to the same material type \cite{Rokhlin2002}, leading to an assembled stiffness matrix with the same dimension as that of a single stiffness matrix. This recursive method, however, is not considered here because we attempt to address a general case with arbitrarily stacked fluid, solid and porous layers. 

The reflection and transmission coefficients for a layered system can be calculated upon substituting the wave conditions in the two half-spaces $0$ and $n+1$ (see Fig. \ref{fig:illustrations}) into the global matrix. Without loss of generality, the two half-spaces are considered to be occupied by the same fluid material, which is the case for air- and water-coupled configurations as commonly used for the testing of layered structures. In the half-space $0$, a monochromatic plane wave is impinged on the $z=0$ surface at an incident angle of $\theta$. Thus, the incident wave has the wavenumber components of
\begin{equation}
    k_{x}=k_\mathrm{L}^0\sin\theta, \, k_{\mathrm{L}z}^0=k_\mathrm{L}^0\cos\theta,
\end{equation}
where $k_\mathrm{L}^0=\omega/c_\mathrm{L}^0$ is the wavenumber in the half-space $0$. As used throughout this paper, the $x$-direction wavenumber component $k_x$ is a common term shared by all waves in the entire structure, which is prescribed by the Snell's law.

As the incident wave comes into the layered medium, part of the wave is reflected back into the half-space $0$, and the rest travels through the layers and is transmitted into the half-space $n+1$. Let us assume the incident wave to have a unit amplitude, then the amplitudes of the reflected and transmitted waves equal to the reflection $R$ and transmission $T$ coefficients of the system. In this case, the amplitude vectors for the half-spaces $0$ and $n+1$ are $\mathbf{A}^0=\left[1,R\right]^\mathrm{T}$ and $\mathbf{A}^{n+1}=\left[T,0\right]^\mathrm{T}$ (note the absence of backward travelling wave in $n+1$). To avoid having an origin at $-\infty$ in the half-space $0$, we place the $z$-coordinate origin of both forward and backward propagating waves at the interface $0$. Then from Eqs. \ref{eq:relation_displacement} and \ref{eq:relation_stress}, we obtain the displacement and stress vectors on the interface $0$ as
\begin{equation}\label{eq:relations_0}
    \mathbf{u}_0^0=\left[k_{\mathrm{L}z}^0(1-R)\right]^\mathrm{T}, \,
    \bm{\sigma}_0^0=\left[i\rho\omega^2(1+R)\right]^\mathrm{T}.
\end{equation}
Similarly, the displacement and stress vectors on the interface $n+1$ can be determined as
\begin{equation}\label{eq:relations_n1}
    \mathbf{u}_{n}^{n+1}=\left[k_{\mathrm{L}z}^0T\right]^\mathrm{T}, \,
    \bm{\sigma}_{n}^{n+1}=\left[i\rho\omega^2T\right]^\mathrm{T},
\end{equation}
where $k_{\mathrm{L}z}^{n+1}=k_{\mathrm{L}z}^0$ is considered. Subsequently, the stiffness matrix relations for the two half-spaces can be obtained from Equations \ref{eq:relations_0} and \ref{eq:relations_n1} as
\begin{equation}\label{eq:relations_both}
    \bm{\sigma}_0^0=\mathbf{K}^0\mathbf{u}_0^0, \,
    \bm{\sigma}_{n}^{n+1}=\mathbf{K}^{n+1}\mathbf{u}_{n}^{n+1},
\end{equation}
with
\begin{equation}
    \mathbf{K}^0=\left[\frac{i\omega Z^0}{\cos\theta} \frac{1+R}{1-R}\right], \,
    \mathbf{K}^{n+1}=\left[\frac{i\omega Z^0}{\cos\theta}\right],
\end{equation}
where $Z^0=\rho^0c_\mathrm{L}^0$ is the acoustic impedance in both half-spaces.

Incorporating Eq. \ref{eq:relations_both} into the global matrix, we have
\begin{widetext}
\begin{equation}\label{eq:global}
    \begin{bmatrix}
    \mathbf{B}_0^{0} & \mathbf{B}_0^{1} & \mathbf{0} & \mathbf{0} & \mathbf{0} & \cdots & \mathbf{0} & \mathbf{0} & \mathbf{0} \\
    \mathbf{C}_0^{0}\mathbf{K}^{0} & \mathbf{C}_0^{1}\mathbf{K}_{11}^{1} & \mathbf{C}_0^{1}\mathbf{K}_{12}^{1} & \mathbf{0} & \mathbf{0} & \cdots & \mathbf{0} & \mathbf{0} & \mathbf{0} \\
    \mathbf{0} & \mathbf{0} & \mathbf{B}_1^{1} & \mathbf{B}_1^{2} & \mathbf{0} & \cdots & \mathbf{0} & \mathbf{0} & \mathbf{0} \\
    \mathbf{0} & \mathbf{C}_1^{1}\mathbf{K}_{21}^{1} & \mathbf{C}_1^{1}\mathbf{K}_{22}^{1} & \mathbf{C}_1^{2}\mathbf{K}_{11}^{2} & \mathbf{C}_1^{2}\mathbf{K}_{12}^{2} & \cdots & \mathbf{0} & \mathbf{0} & \mathbf{0} \\
    \vdots & \vdots & \vdots & \vdots & \vdots & \ddots & \vdots & \vdots & \vdots \\
    \mathbf{0} & \mathbf{0} & \mathbf{0} & \mathbf{0} & \mathbf{0} & \cdots & \mathbf{0} & \mathbf{B}_{n}^{n} & \mathbf{B}_{n}^{n+1} \\
    \mathbf{0} & \mathbf{0} & \mathbf{0} & \mathbf{0} & \mathbf{0} & \cdots & \mathbf{C}_n^{n}\mathbf{K}_{21}^{n} & \mathbf{C}_n^{n}\mathbf{K}_{22}^{n} & \mathbf{C}_n^{n+1}\mathbf{K}^{n+1}
    \end{bmatrix}
    \begin{bmatrix}
        \mathbf{u}_{0}^{0} \\ \mathbf{u}_{0}^{1} \\ \mathbf{u}_{1}^{1} \\ \mathbf{u}_{1}^{2} \\ \mathbf{u}_{2}^{2} \\ \vdots \\ \mathbf{u}_{n-1}^{n} \\ \mathbf{u}_{n}^{n} \\ \mathbf{u}_{n}^{n+1}
    \end{bmatrix} = \mathbf{V}\mathbf{U} = \mathbf{0},
\end{equation}
\end{widetext}
where the global matrix $\mathbf{V}$ is a square matrix with an even size, and it has only one unknown variable of $R$. For the equation to have nontrivial solutions, the determinant of $\mathbf{V}$ must vanish. We can observe from $\mathbf{V}$ that its first column has only two nonzero elements on the first and second rows, so its determinant can be expressed as
\begin{equation}\label{eq:det}
    \det\mathbf{V}=\det\mathbf{V}_{1,1}-\left(\frac{i\omega Z^0}{\cos\theta} \frac{1+R}{1-R}\right)\det\mathbf{V}_{2,1}=0,
\end{equation}
where the values of the two nonzero elements have been incorporated. $\mathbf{V}_{i,j}$ is the submatrix formed by removing the $i$-th row and $j$-th column of $\mathbf{V}$. Equation \ref{eq:det} produces the result for the reflection coefficient as
\begin{equation}\label{eq:R}
    R=\frac{Z^\mathrm{eff}-Z^0}{Z^\mathrm{eff}+Z^0},
\end{equation}
with
\begin{equation}\label{eq:Z_surface}
    Z^\mathrm{eff}=\frac{\cos\theta}{i\omega}\frac{\det\mathbf{V}_{1,1}}{\det\mathbf{V}_{2,1}}.
\end{equation}

To calculate the transmission coefficient $T$, we consider the displacement relation between the two half-spaces, obtained from Eqs. \ref{eq:relations_0} and \ref{eq:relations_n1} as $T\mathbf{u}_0^0+(R-1)\mathbf{u}_n^{n+1}=\mathbf{0}$. Replacing the first row of $\mathbf{V}$ in Eq. \ref{eq:global} with this relation and evaluating the determinant of the resulting matrix, we arrive at the expression for the transmission coefficient
\begin{equation}\label{eq:T}
    T = (R-1)\frac{\det\mathbf{\mathbf{V}}_\mathrm{1,end}}{\det\mathbf{\mathbf{V}}_{1,1}},
\end{equation}
where `end' represents the last column of $\mathbf{V}$.

The calculated $R$ and $T$ are complex numbers, carrying both amplitude and phase information of the reflected and transmitted waves. We point out that we can readily obtain the three acoustic indicators commonly used in sound-absorbing applications, with surface impedance given by Eq. \ref{eq:Z_surface}, absorption coefficient by $1-|R|^2$ and transmission loss by $-10\log|T^2|$ \cite{Allard2009}. Though it will not be discussed in detail here, we emphasise that above formulations can be easily adapted to other boundary conditions, such as those bounded by an impervious hard wall on one side and those by solids on both sides.

\subsection{Relation to transfer matrix method and stability}

The stiffness matrix method is closely related to the well-established transfer matrix method \cite{Brouard1995,Allard2009,Jocker2004}. To demonstrate this relationship, we reorganise the displacements and stresses in Eqs. \ref{eq:relation_displacement} and \ref{eq:relation_stress} to the two layer interfaces, yielding
\begin{align}
    \begin{bmatrix} \mathbf{u}_{i-1} \\ \bm{\sigma}_{i-1} \end{bmatrix}^i = 
    \begin{bmatrix} \mathbf{D}^+ & \mathbf{D}^-\mathbf{E} \\ \mathbf{S}^+ & \mathbf{S}^-\mathbf{E} \end{bmatrix}^i
    \begin{bmatrix} \mathbf{A}^+ \\ \mathbf{A}^- \end{bmatrix}^i, \label{eq:transfer1}\\
    \begin{bmatrix} \mathbf{u}_{i} \\ \bm{\sigma}_{i} \end{bmatrix}^i = 
    \begin{bmatrix} \mathbf{D}^+\mathbf{E} & \mathbf{D}^- \\ \mathbf{S}^+\mathbf{E} & \mathbf{S}^- \end{bmatrix}^i
    \begin{bmatrix} \mathbf{A}^+ \\ \mathbf{A}^- \end{bmatrix}^i. \label{eq:transfer2}
\end{align}
Inverting Eq. \ref{eq:transfer2} and substituting the resulting amplitude vector into Eq. \ref{eq:transfer1} leads to a transfer matrix relation
\begin{align}\small
    \begin{bmatrix} \mathbf{u}_{i-1} \\ \bm{\sigma}_{i-1} \end{bmatrix}^i
    &=\begin{bmatrix} \mathbf{D}^+ & \mathbf{D}^-\mathbf{E} \\ \mathbf{S}^+ & \mathbf{S}^-\mathbf{E} \end{bmatrix}^i
    \left(\begin{bmatrix} \mathbf{D}^+\mathbf{E} & \mathbf{D}^- \\ \mathbf{S}^+\mathbf{E} & \mathbf{S}^- \end{bmatrix}^i\right)^{-1}
    \begin{bmatrix} \mathbf{u}_{i} \\ \bm{\sigma}_{i} \end{bmatrix}^i \\
    &=\mathbf{T}^i \begin{bmatrix} \mathbf{u}_{i} \\ \bm{\sigma}_{i} \end{bmatrix}^i,\nonumber
\end{align}
which relates the displacements and stresses on the interface $i-1$ to those on the interface $i$ of the layer $i$. The elements of the transfer matrix $\mathbf{T}$ can be expressed by those of the stiffness matrix $\mathbf{K}$ (and vice versa) \cite{Rokhlin2002}; the expressions are not provided here because they are numerically unstable. Using the boundary conditions in Eqs. \ref{eq:bc_displacement} and \ref{eq:bc_stress}, the transfer matrix relation is assembled into a global matrix form in order to solve the wave reflection and transmission problem in the layered system in the same fashion as presented above \cite{Brouard1995,Allard2009}.

\begin{figure}
    \centering
    \includegraphics[width=\linewidth]{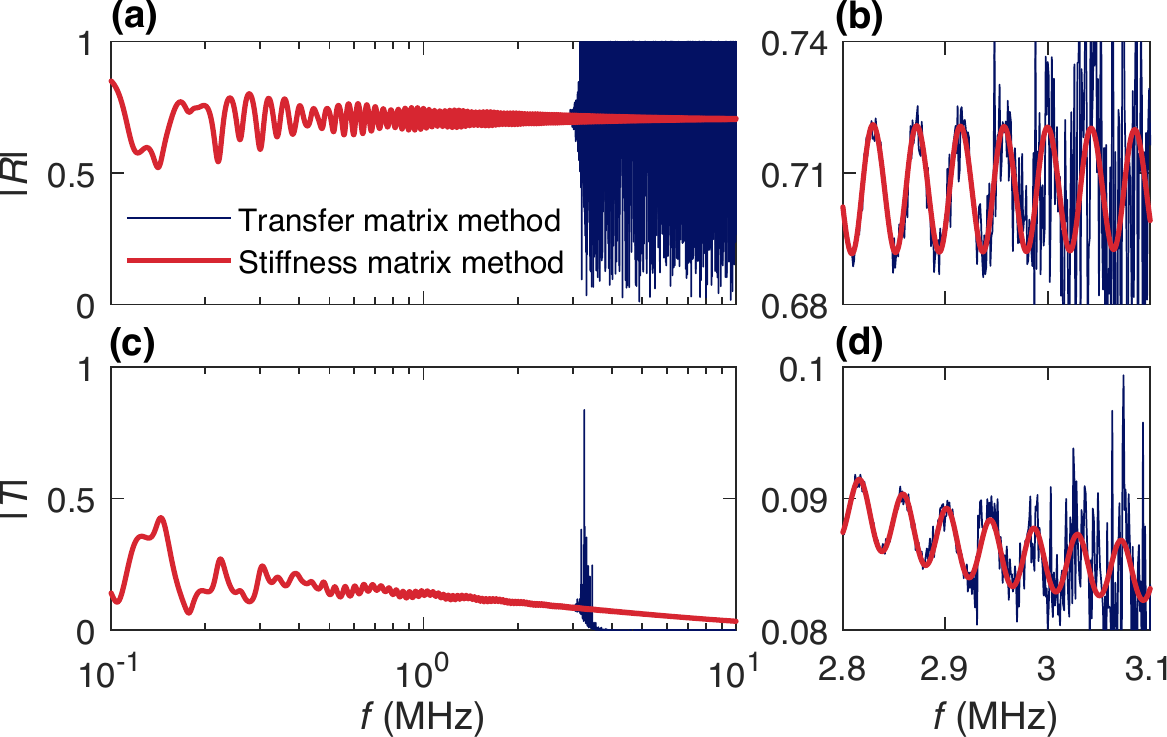}
    \caption{Comparison of the transfer matrix and stiffness matrix methods for wave reflection and transmission in a porous layer at an incident angle of $\theta=40^\circ$ (beyond the critical angle of the fast longitudinal wave). The porous layer is made of sintered glass beads saturated with water and the entire layer is bounded by water \cite{Jocker2009}, and this calculation case will be studied in detail in Fig. \ref{fig:glass}. (a) and (c) Amplitudes of the reflection and transmission coefficients, $|R|$ and $|T|$, obtained from the two methods. (b) and (d) Respective zoomed-in plots, highlighting the unstable blow-ups of the transfer matrix method.}
    \label{fig:stability}
\end{figure}

As mentioned, the stiffness matrix formulation brings the benefit of intrinsic stability. To compare it with transfer matrix method, we have run a wide variety of calculation cases with the layered system containing different numbers of layers of different material types. The stiffness matrix method is found to be numerically stable in all cases; while the transfer matrix method delivers practically identical results when numerically-stable, it does suffer from instability under certain conditions, even when assembled in the global matrix configuration. This might seem contradicting to the discussions in \cite{Lowe1995} and in the Introduction - the transfer matrix method is supposed to be free from instability in global matrix formulation. This is found to be indeed true for the cases involving fluid and/or solid layers only. However, instability does arise for porous layers, potentially due to the existence of slow waves (slower speed than waves in fluid). A prominent example case is provided in Figure \ref{fig:stability} for a porous layer, showing unstable blow-ups at high frequencies beyond the first critical angle. Another unstable case arises at the $f\to0$ limit where instability leads to a non-unity transmission coefficient (supposed to be unity because the layered structure is transparent to the transmitting wave). This latter case is less important so is not plotted here. Our stiffness matrix method does not suffer from these issues and is thus more advantageous.

\section{Results and experimental validation}\label{sec:results}

\begin{figure}
    \centering
    \includegraphics[width=\linewidth]{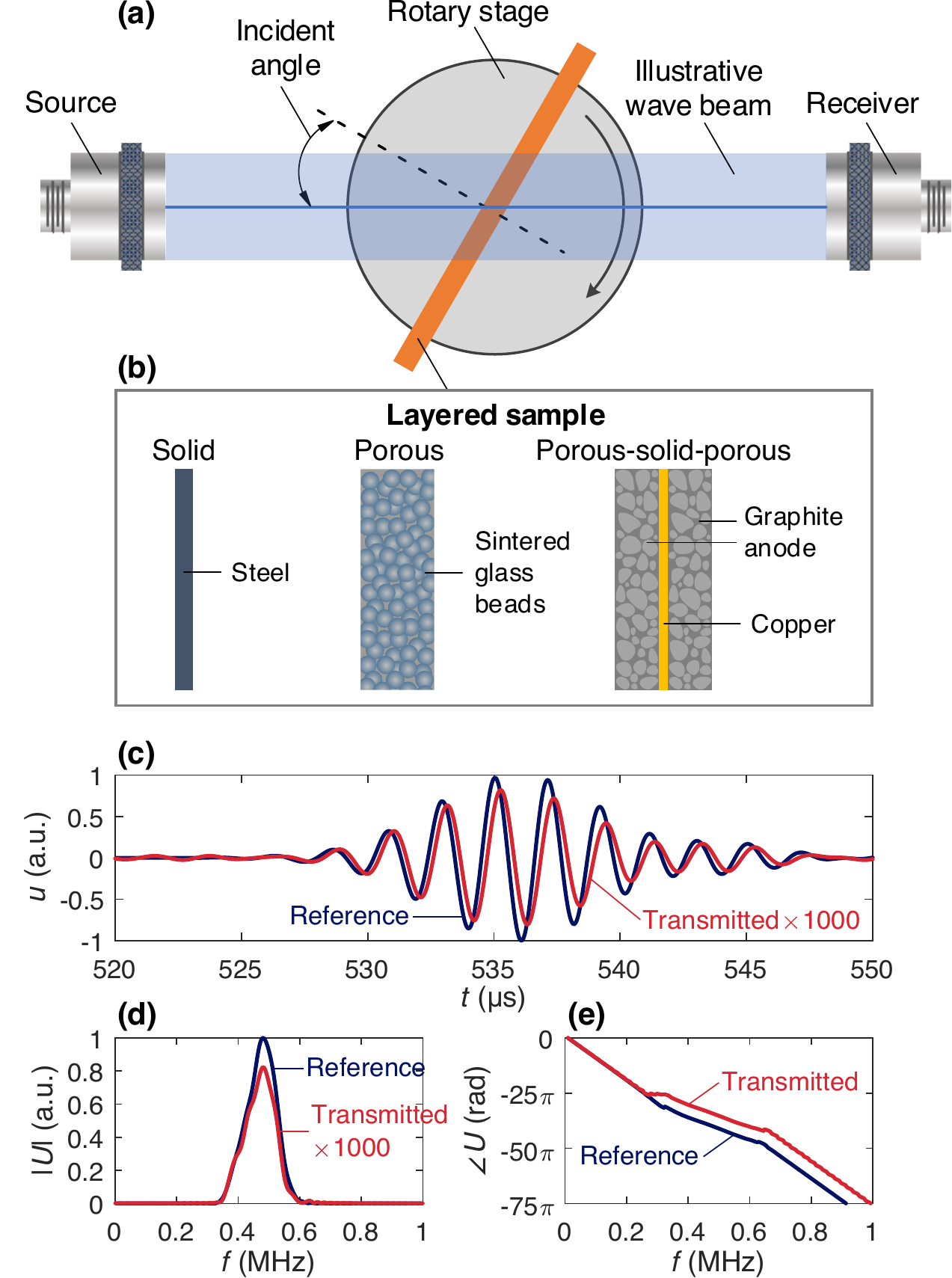}
    \caption{Experimental setup, samples and example signals. (a) Experimental setup with a transducer generating wave into a layered sample and another transducer receiving the transmitted wave. The sample is mounted onto a rotary stage to adjust the angle of incidence. (b) Three validation samples, with the first being a solid layer, the second a porous layer and the third a porous-solid-porous layer. (c) Example reference signal recorded when the sample is absent and transmitted signal when the wave has a $22.5^\circ$ incident angle to the $\SI{50}{\micro\metre}$ steel sample. Each signal is the average of 256 recordings acquired using a pair of 0.5 MHz transducers. (d) Amplitude spectra of the two example signals. (e) Respective phase spectra.}
    \label{fig:setup}
\end{figure}

Having fully established the stiffness matrix method, we present wave propagation results predicted by it (referred to as `theory') and compare them with experimental measurements for a range of layered media. We employ the experimental setup sketched in Fig. \ref{fig:setup}(a). The sample is placed between two ultrasonic transducers, with the source transducer generating wave into the coupling fluid and the receiver transducer recording the wave that travelled through the fluid-sample-fluid path. The sample is mounted vertically onto a stepper motor-driven rotary stage and the angle $\theta$ between the incident wave and the sample surface is thus automatically controlled. The two transducers are each fixed onto a kinematic mount, through which they are adjusted to have their axes aligned and their active surfaces parallel to the sample surface at the normal incidence of $\theta=0^\circ$.

We consider three samples of gradually increasing complexity as illustrated in Fig. \ref{fig:setup}(b). To experimentally measure the transmission coefficient of each sample, a reference signal $u_\mathrm{r}(t)$ is first acquired when the sample is absent, and then the transmitted signal $u_\mathrm{t}(t)$ through the sample is recorded at a desired incident angle $\theta$. Example reference signal and transmitted signal through the solid steel sample at $22.5^\circ$ are provided in Fig. \ref{fig:setup}(c), which are acquired using a pair of 0.5 MHz transducers. The two signals are subsequently transformed into the frequency domain to obtain the spectra $U_\mathrm{r}(f)$ and $U_\mathrm{t}(f)$; see Fig. \ref{fig:setup}(d) and (e) for the amplitude and phase spectra of the example signals. The frequency-dependent transmission coefficient at $\theta$ is calculated by $T(f)=U_\mathrm{t}(f)/U_\mathrm{r}(f)e^{-ik_\mathrm{f}d\cos\theta}$, with the exponential term accounting for the propagation across the sample thickness $d$ in the coupling fluid (with wave number $k_\mathrm{f}\cos\theta$ in the propagation direction) when the sample is absent. The transmission coefficient is mostly small especially in an air-coupled setting, and the transmitted signal is thus very weak. To reach a signal-to-noise ratio of around 30 dB for the transmitted signal, the voltage of the excitation (5-cycle Hann-windowed toneburst) to the source transducer is maintained at around 100 V and the recorded signal by the receiver transducer is pre-amplified by 40 dB. In addition, each signal recording is the average of 256 signal firings, as a commonly-employed technique to suppress electronic noises.

\subsection{Single solid layer}

\begin{figure}
    \centering
    \includegraphics[width=\linewidth]{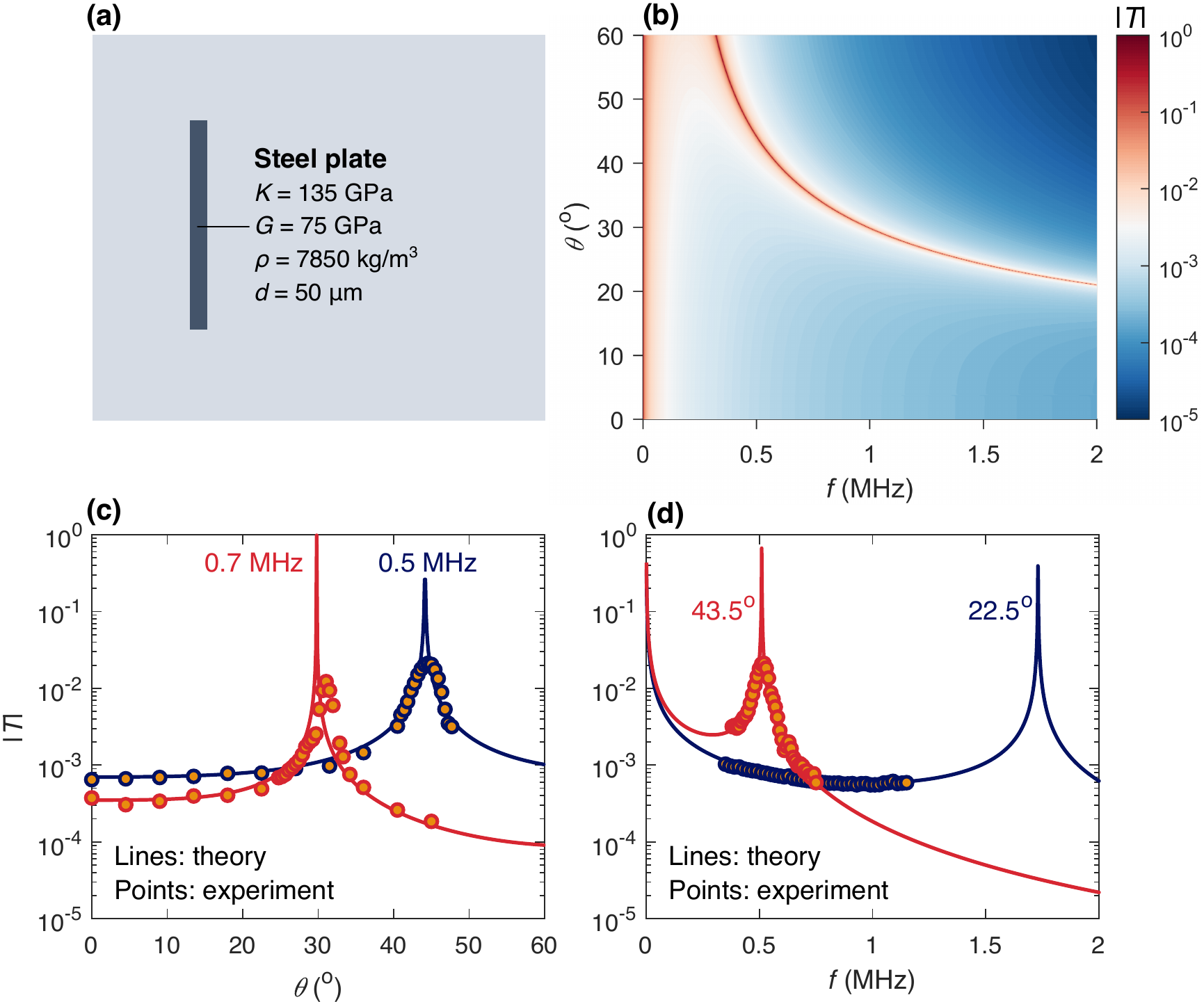}
    \caption{Wave transmission through a solid layer. (a) The properties of the thin steel sample with a thickness of \SI{50}{\micro\metre}. (b) The amplitude of theoretically predicted transmission coefficient as a function of frequency $f$ and incident angle $\theta$. (c) Comparison of theoretical and experimental results at the frequencies of 0.5 and 0.7 MHz. (d) Similar comparison at the incident angles of 22.5$^\circ$ and 43.5$^\circ$.}
    \label{fig:steel}
\end{figure}

We start with a simple case of a solid layer submerged in air at 20 $^\circ$C room temperature. The properties of the sample is given in Fig. \ref{fig:steel}(a), and those of air are $K=1.4\times10^5$ Pa, $\rho=1.3$ kg/m\textsuperscript{3} and $\eta=1.8\times10^{-5}$ Pa$\cdot$s. This case is an important first step for understanding wave propagation in multilayered media, such as battery electrodes that have a thin solid layer in the middle as discussed later.

The theoretical transmission coefficient amplitude $|T|$ is provided in Fig. \ref{fig:steel}(b) as a function of frequency $f$ and incident angle $\theta$. Since the amplitude spans over a few orders of magnitude, a logarithmic colour scale is used in the plot for better visualisation. At low frequencies where the wavelength is substantially larger than the sample thickness, the transmission coefficient in all directions approaches unity, meaning that nearly all energy is transmitted through the layer as if the layer is not present. The transmission coefficient is mostly very small at other frequencies, but it tends to reach unity again at large incident angles beyond the two critical angles of $3.5^\circ$ and $6.2^\circ$. This large-angle, highly-transmissible region arises due to the excitation of the fundamental anti-symmetric guided wave (A\textsubscript{0} mode) in the thin layer as detailed in our prior work \cite{Almeida2023}. Physically, the wave resonates in the layer at the presence of the A\textsubscript{0} mode, and it causes the majority of the energy to travel through the layer to the other side.

The theoretical prediction is corroborated by the experimental results in Fig. \ref{fig:steel}(c) and (d) that are collected using three pairs of air-coupled transducers with the frequencies of 0.5, 0.7 and 1.0 MHz. The two figure plots display respectively the angle dependence of the transmission coefficient amplitude at $f=0.5$ and 0.7 MHz and the frequency dependence at $\theta=22.5^\circ$ and $43.5^\circ$. Both plots show very good agreement between the theory and the experiment. Prominently, in comparison to the experiment, the theory has accurately predicted the fine details at, and around, the highly-transmissible A\textsubscript{0} peaks. We have also carried out experiments on 250 and \SI{500}{\micro\metre} thick samples and the results show similar theory-experiment agreement and transmission characteristics.

\subsection{Single porous layer}

\begin{figure}
    \centering
    \includegraphics[width=\linewidth]{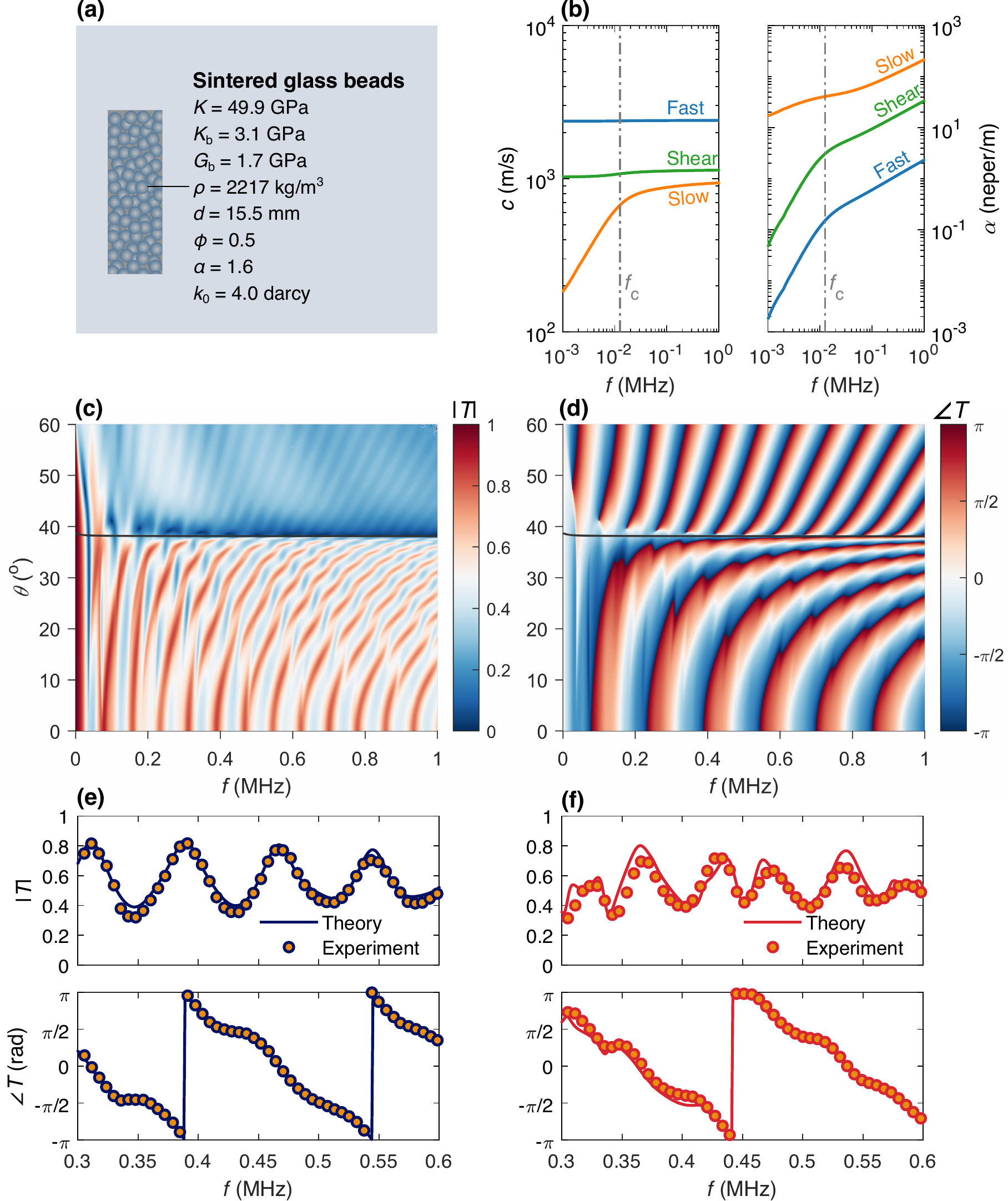}
    \caption{Wave transmission through a porous layer. (a) Properties of the porous material of sintered glass beads saturated with water \cite{Jocker2009}. (b) Wave speed (left) and attenuation coefficient (right) of the fast and slow longitudinal waves and the shear wave in the porous material. The vertical dash-dotted line represents the viscous characteristic frequency $f_\mathrm{c}$. (c) and (d) Amplitude and phase maps of theoretically predicted transmission coefficient against frequency $f$ and incident angle $\theta$. The solid line in the maps denotes the first critical angle for the fast longitudinal wave. (e) Comparison of amplitude (top) and phase (bottom) between the theoretical and experimental results at the normal incidence of $\theta=0^\circ$. (f) Similar comparison at the incident angle of $\theta=18^\circ$. The experimental points in (e) and (f) are taken from Jocker et al. \cite{Jocker2009}.}
    \label{fig:glass}
\end{figure}

Going further, we analyse wave transmission through a porous layer of the same granular microstructure as active battery electrodes. We consider a well-studied porous material made of sintered glass beads \cite{Johnson1982,Liu1990,Hu2008,Jocker2009} as its microstructure characteristics are highly controllable. We use the parameters and experimental results of the sample S3 by Jocker et al. \cite{Jocker2009}. The sample has the parameters in Fig. \ref{fig:glass}(a) and is submerged in water of $K=2.2$ GPa, $\rho=1000$ kg/m\textsuperscript{3} and $\eta=0.001$ Pa$\cdot$s. Figure \ref{fig:glass}(b) gives the wave speed and attenuation coefficient of the three Biot waves in the porous medium, namely the fast and slow longitudinal waves and the shear wave. As affected by the porous network (porosity, tortuosity, permeability etc), the two faster waves travel more slowly than in pure glass (5850 and 3250 m/s \cite{Johnson1982a}) and the slow wave is even slower than in the fluid (1483 m/s). Also, the three waves exhibit different levels of attenuation, increasing around an order of magnitude each from the fast longitudinal through the shear to the slow longitudinal waves. The three waves show distinctive wave speed and attenuation characteristics in the low-frequency viscous and high-frequency inertial regimes separated by the viscous characteristic frequency $f_\mathrm{c}=0.013$ MHz. In the viscous regime, the slow wave demonstrates more pronounced wave speed dispersion but smaller increase of attenuation with frequency than the other two waves. By contrast, all three waves tend to have constant speeds and the same power dependence of attenuation on frequency in the inertial regime.

Our theoretical prediction of the transmission coefficient through the sample is given in Fig. \ref{fig:glass}(c) and (d), showing respectively the amplitude and phase maps against frequency $f$ and incident angle $\theta$. The transmission is small beyond the critical angle of the fast longitudinal wave, which is annotated as the solid line in the maps. Below this critical angle, the transmission coefficient exhibits a cyclic behaviour, which is particularly obvious as we look along the horizontal line at the normal incidence of $\theta=0^\circ$ in the amplitude map. This cyclic behaviour arises because of the resonances of the three waves, predominately the fast longitudinal wave, reverberating between the two boundaries of the porous layer. The resonances also cause small phase fluctuations as can be seen in the phase map.

Figure \ref{fig:glass}(c) and (d) compare our theoretical prediction with experimental results \cite{Jocker2009} at the incident angles of $\theta=0^\circ$ and $\theta=18^\circ$, respectively. In both cases, the theory agrees remarkably well with the experimental points in both amplitude and phase, building our confidence in using the theory to describe multilayered porous electrodes in what follows.

\begin{figure}[t]
    \centering
    \includegraphics[width=\linewidth]{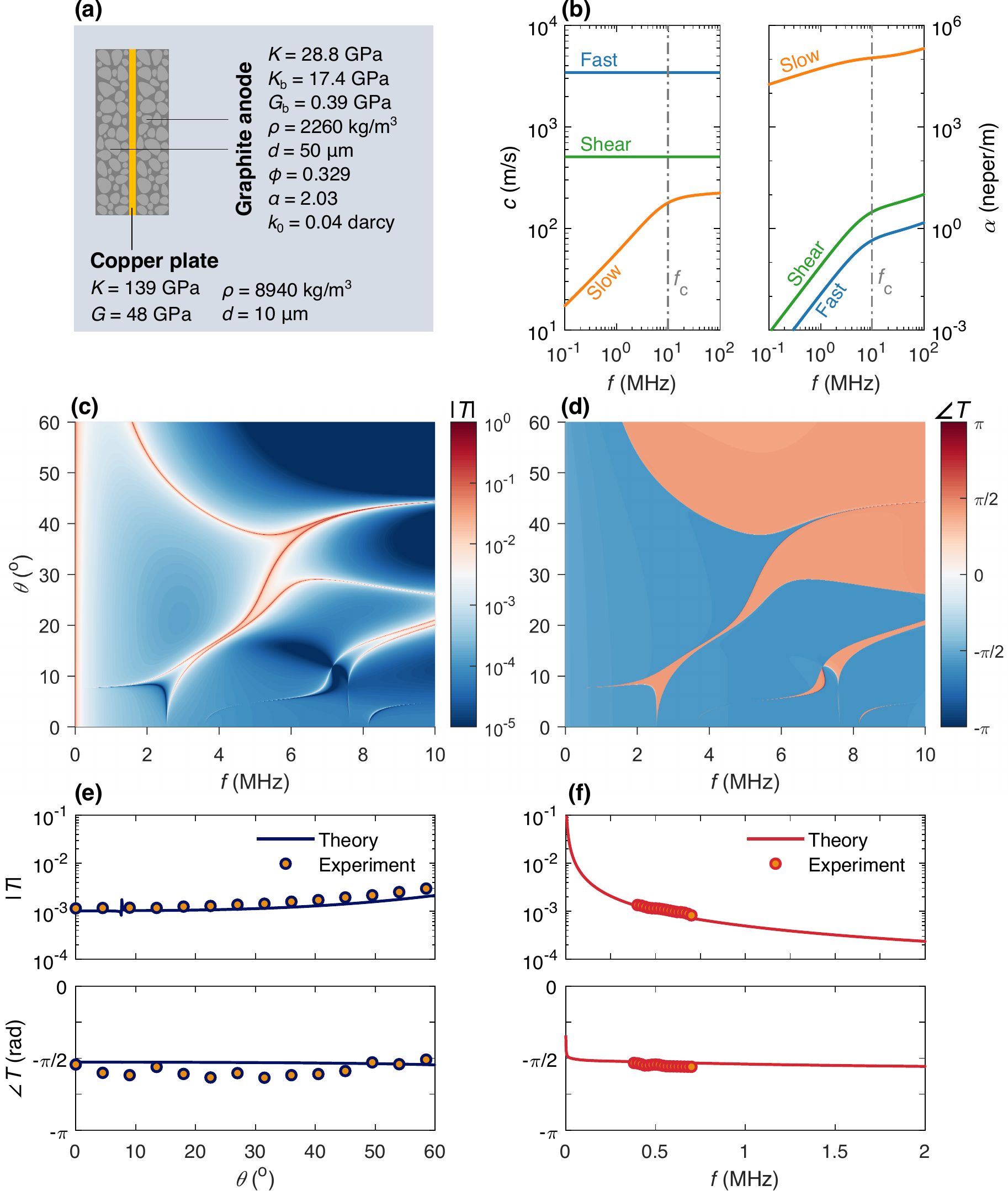}
    \caption{Wave transmission through a porous-solid-porous anode. (a) Properties of the multilayered anode \cite{Ecker2015,Davies2017,Gold2017} in an air-coupled setting. (b) Wave speed (left) and attenuation coefficient (right) of the three Biot waves in the graphite anode coating material, with the viscous characteristic frequency $f_\mathrm{c}$ annotated as the vertical dash-dotted line. (c) and (d) Amplitude and phase maps of theoretically predicted transmission coefficient against frequency $f$ and incident angle $\theta$. (e) Angular dependence of amplitude (top) and phase (bottom) at 0.5 MHz, comparing the theoretical and experimental results. (f) Frequency dependence at the normal incidence of $\theta=0^\circ$.}
    \label{fig:anode}
\end{figure}

\subsection{Porous-solid-porous anode}

Now we examine wave transmission through a porous-solid-porous battery anode. We focus on the most-widely anode material with two $\SI{50}{\micro\metre}$ active layers coated on both sides of a $\SI{10}{\micro\metre}$ copper film. The active layer has a granular porous microstructure with graphite particles joined together by binder materials. The active material has been well characterised using advanced techniques \cite{Ecker2015,Davies2017,Gold2017} and the relevant parameters are summarised in Fig. \ref{fig:anode}(a). The three Biot waves in the active material are detailed in Fig. \ref{fig:anode}(b), showing similar speed and attenuation profiles to those in sintered glass beads in Fig. \ref{fig:glass}(b). The use of airborne ultrasound in this case and the micron-level pore structure lead the slow longitudinal wave to have exceptionally small speed and excessively large attenuation. Thus, in comparison to the other two waves, the slow wave has minimal contribution to the transmitted wave. In addition, the large viscous characteristic frequency of $f_\mathrm{c}=9.75$ MHz means that we have to focus in the viscous regime because the inertial regime can barely be reached with an air-coupled setup (generally limited to $\leq5$ MHz).

Figure \ref{fig:anode}(c) and (d) display respectively the theoretical amplitude and phase maps of the transmission coefficient through the anode. As in a single solid or porous layer, the wave perceives the anode as transparent irrespective of the incident angle in the low-frequency, long-wavelength range. At higher frequencies, the wave tunnels through the anode when the frequency and incident angle are suitably combined to excite guided wave modes in individual layers or in the whole anode, evidenced also by the $\pi$ phase jump in the phase map. The highly-transmissible tunneling is reminiscent of the above single solid layer but shows many more complex features due to the multilayered nature of the anode.

The theoretical results are evaluated against experimental measurements in Fig. \ref{fig:anode}(e) and (f). The theory matches very well with the experiment, particularly in the transmission amplitude, from both the angle- and frequency-dependent results. The agreement is less satisfactory for the phase results in Fig. \ref{fig:anode}(e) because of the difficulty in obtaining accurate phase information from noisy experimental signals. Overall, the evaluation results highlight the very good applicability of the theory to describe wave transmission through complex multilayered electrodes involving micron-level porous structures. 

\section{Summary and outlook}\label{sec:summary}

In summary, we have developed a general stiffness matrix method for modelling wave propagation in multilayered media with arbitrary numbers and combinations of fluid, solid and porous layers. With individual layers described by stiffness matrices, and layer interfaces defined by boundary conditions, the proposed method assembles a global system of equations to solve the reflection and transmission of oblique waves in the layered system. The method is intrinsically and unconditionally stable and is thus more advantageous over existing transfer matrix-based methods, which have been demonstrated to show instability in certain configurations. Experimental validations have proved the validity of the proposed method over a range of layered cases, ranging from single solid and porous layers to a porous-solid-porous anode.

The presented work is a powerful modelling tool for developing new methods to quantify the properties of layered media. A particularly exciting possibility is the quantification of the porous parameters (e.g. porosity, tortuosity and permeability) in Li-ion battery electrodes, which we have already analysed in this paper tentatively. These electrode parameters are key performance determinants of the batteries, hence non-destructive evaluation capabilities, potentially offered by wave-based methods, could bring considerable benefits via closed-loop control and quality assurance during battery manufacturing.

\begin{acknowledgments}
We acknowledge Prof Kirill Horoshenkov of Sheffield University for insightful discussions on poroelasticity, Antonio De Sanctis of Imperial College and Peiyao Huang of Cambridge University for assistance in experimental setup. B.L. gratefully thanks the generous support from the Imperial College Research Fellowship Scheme, and M.H. appreciates the funding by the Imperial College Non-Destructive Evaluation Group.
\end{acknowledgments}

\bibliography{references}

\begin{thebibliography}{51}%
\makeatletter
\providecommand \@ifxundefined [1]{%
 \@ifx{#1\undefined}
}%
\providecommand \@ifnum [1]{%
 \ifnum #1\expandafter \@firstoftwo
 \else \expandafter \@secondoftwo
 \fi
}%
\providecommand \@ifx [1]{%
 \ifx #1\expandafter \@firstoftwo
 \else \expandafter \@secondoftwo
 \fi
}%
\providecommand \natexlab [1]{#1}%
\providecommand \enquote  [1]{``#1''}%
\providecommand \bibnamefont  [1]{#1}%
\providecommand \bibfnamefont [1]{#1}%
\providecommand \citenamefont [1]{#1}%
\providecommand \href@noop [0]{\@secondoftwo}%
\providecommand \href [0]{\begingroup \@sanitize@url \@href}%
\providecommand \@href[1]{\@@startlink{#1}\@@href}%
\providecommand \@@href[1]{\endgroup#1\@@endlink}%
\providecommand \@sanitize@url [0]{\catcode `\\12\catcode `\$12\catcode
  `\&12\catcode `\#12\catcode `\^12\catcode `\_12\catcode `\%12\relax}%
\providecommand \@@startlink[1]{}%
\providecommand \@@endlink[0]{}%
\providecommand \url  [0]{\begingroup\@sanitize@url \@url }%
\providecommand \@url [1]{\endgroup\@href {#1}{\urlprefix }}%
\providecommand \urlprefix  [0]{URL }%
\providecommand \Eprint [0]{\href }%
\providecommand \doibase [0]{https://doi.org/}%
\providecommand \selectlanguage [0]{\@gobble}%
\providecommand \bibinfo  [0]{\@secondoftwo}%
\providecommand \bibfield  [0]{\@secondoftwo}%
\providecommand \translation [1]{[#1]}%
\providecommand \BibitemOpen [0]{}%
\providecommand \bibitemStop [0]{}%
\providecommand \bibitemNoStop [0]{.\EOS\space}%
\providecommand \EOS [0]{\spacefactor3000\relax}%
\providecommand \BibitemShut  [1]{\csname bibitem#1\endcsname}%
\let\auto@bib@innerbib\@empty
\bibitem [{\citenamefont {H{\"{a}}gglund}\ \emph {et~al.}(2009)\citenamefont
  {H{\"{a}}gglund}, \citenamefont {Martinsson},\ and\ \citenamefont
  {Carlson}}]{Hagglund2009}%
  \BibitemOpen
  \bibfield  {author} {\bibinfo {author} {\bibfnamefont {F.}~\bibnamefont
  {H{\"{a}}gglund}}, \bibinfo {author} {\bibfnamefont {J.}~\bibnamefont
  {Martinsson}},\ and\ \bibinfo {author} {\bibfnamefont {J.~E.}\ \bibnamefont
  {Carlson}},\ }\bibfield  {title} {\bibinfo {title} {{Model-based estimation
  of thin multi-layered media using ultrasonic measurements}},\ }\href
  {https://doi.org/10.1109/TUFFC.2009.1233} {\bibfield  {journal} {\bibinfo
  {journal} {IEEE Transactions on Ultrasonics, Ferroelectrics, and Frequency
  Control}\ }\textbf {\bibinfo {volume} {56}},\ \bibinfo {pages} {1689}
  (\bibinfo {year} {2009})}\BibitemShut {NoStop}%
\bibitem [{\citenamefont {Lerch}(2021)}]{Lerch2021}%
  \BibitemOpen
  \bibfield  {author} {\bibinfo {author} {\bibfnamefont {T.~P.}\ \bibnamefont
  {Lerch}},\ }\bibfield  {title} {\bibinfo {title} {{Ultrasonic thickness
  measurement of nonporous membranes with long wavelengths}},\ }\href
  {https://doi.org/10.1016/j.ultras.2021.106370} {\bibfield  {journal}
  {\bibinfo  {journal} {Ultrasonics}\ }\textbf {\bibinfo {volume} {113}},\
  \bibinfo {pages} {106370} (\bibinfo {year} {2021})}\BibitemShut {NoStop}%
\bibitem [{\citenamefont {Smith}\ \emph {et~al.}(2018)\citenamefont {Smith},
  \citenamefont {Nelson}, \citenamefont {Mienczakowski},\ and\ \citenamefont
  {Wilcox}}]{Smith2018}%
  \BibitemOpen
  \bibfield  {author} {\bibinfo {author} {\bibfnamefont {R.~A.}\ \bibnamefont
  {Smith}}, \bibinfo {author} {\bibfnamefont {L.~J.}\ \bibnamefont {Nelson}},
  \bibinfo {author} {\bibfnamefont {M.~J.}\ \bibnamefont {Mienczakowski}},\
  and\ \bibinfo {author} {\bibfnamefont {P.~D.}\ \bibnamefont {Wilcox}},\
  }\bibfield  {title} {\bibinfo {title} {{Ultrasonic Analytic-Signal Responses
  from Polymer-Matrix Composite Laminates}},\ }\href
  {https://doi.org/10.1109/TUFFC.2017.2774776} {\bibfield  {journal} {\bibinfo
  {journal} {IEEE Transactions on Ultrasonics, Ferroelectrics, and Frequency
  Control}\ }\textbf {\bibinfo {volume} {65}},\ \bibinfo {pages} {231}
  (\bibinfo {year} {2018})}\BibitemShut {NoStop}%
\bibitem [{\citenamefont {Yang}\ \emph {et~al.}(2021)\citenamefont {Yang},
  \citenamefont {Verboven}, \citenamefont {Ju},\ and\ \citenamefont
  {Kersemans}}]{Yang2021}%
  \BibitemOpen
  \bibfield  {author} {\bibinfo {author} {\bibfnamefont {X.}~\bibnamefont
  {Yang}}, \bibinfo {author} {\bibfnamefont {E.}~\bibnamefont {Verboven}},
  \bibinfo {author} {\bibfnamefont {B.-f.}\ \bibnamefont {Ju}},\ and\ \bibinfo
  {author} {\bibfnamefont {M.}~\bibnamefont {Kersemans}},\ }\bibfield  {title}
  {\bibinfo {title} {{Comparative study of ultrasonic techniques for
  reconstructing the multilayer structure of composites}},\ }\href
  {https://doi.org/10.1016/j.ndteint.2021.102460} {\bibfield  {journal}
  {\bibinfo  {journal} {NDT \& E International}\ }\textbf {\bibinfo {volume}
  {121}},\ \bibinfo {pages} {102460} (\bibinfo {year} {2021})}\BibitemShut
  {NoStop}%
\bibitem [{\citenamefont {Copley}\ \emph {et~al.}(2021)\citenamefont {Copley},
  \citenamefont {Cumming}, \citenamefont {Wu},\ and\ \citenamefont
  {Dwyer-Joyce}}]{Copley2021}%
  \BibitemOpen
  \bibfield  {author} {\bibinfo {author} {\bibfnamefont {R.~J.}\ \bibnamefont
  {Copley}}, \bibinfo {author} {\bibfnamefont {D.}~\bibnamefont {Cumming}},
  \bibinfo {author} {\bibfnamefont {Y.}~\bibnamefont {Wu}},\ and\ \bibinfo
  {author} {\bibfnamefont {R.~S.}\ \bibnamefont {Dwyer-Joyce}},\ }\bibfield
  {title} {\bibinfo {title} {{Measurements and modelling of the response of an
  ultrasonic pulse to a lithium-ion battery as a precursor for state of charge
  estimation}},\ }\href {https://doi.org/10.1016/j.est.2021.102406} {\bibfield
  {journal} {\bibinfo  {journal} {Journal of Energy Storage}\ }\textbf
  {\bibinfo {volume} {36}},\ \bibinfo {pages} {102406} (\bibinfo {year}
  {2021})}\BibitemShut {NoStop}%
\bibitem [{\citenamefont {Huang}\ \emph {et~al.}(2022)\citenamefont {Huang},
  \citenamefont {Kirkaldy}, \citenamefont {Zhao}, \citenamefont {Patel},
  \citenamefont {Cegla},\ and\ \citenamefont {Lan}}]{Huang2022b}%
  \BibitemOpen
  \bibfield  {author} {\bibinfo {author} {\bibfnamefont {M.}~\bibnamefont
  {Huang}}, \bibinfo {author} {\bibfnamefont {N.~D.}\ \bibnamefont {Kirkaldy}},
  \bibinfo {author} {\bibfnamefont {Y.}~\bibnamefont {Zhao}}, \bibinfo {author}
  {\bibfnamefont {Y.}~\bibnamefont {Patel}}, \bibinfo {author} {\bibfnamefont
  {F.}~\bibnamefont {Cegla}},\ and\ \bibinfo {author} {\bibfnamefont
  {B.}~\bibnamefont {Lan}},\ }\bibfield  {title} {\bibinfo {title}
  {{Quantitative characterisation of the layered structure within lithium-ion
  batteries using ultrasonic resonance}},\ }\href
  {https://doi.org/10.1016/j.est.2022.104585} {\bibfield  {journal} {\bibinfo
  {journal} {Journal of Energy Storage}\ }\textbf {\bibinfo {volume} {50}},\
  \bibinfo {pages} {104585} (\bibinfo {year} {2022})}\BibitemShut {NoStop}%
\bibitem [{\citenamefont {Thomson}(1950)}]{Thomson1950}%
  \BibitemOpen
  \bibfield  {author} {\bibinfo {author} {\bibfnamefont {W.~T.}\ \bibnamefont
  {Thomson}},\ }\bibfield  {title} {\bibinfo {title} {{Transmission of elastic
  waves through a stratified solid medium}},\ }\href
  {https://doi.org/10.1063/1.1699629} {\bibfield  {journal} {\bibinfo
  {journal} {Journal of Applied Physics}\ }\textbf {\bibinfo {volume} {21}},\
  \bibinfo {pages} {89} (\bibinfo {year} {1950})}\BibitemShut {NoStop}%
\bibitem [{\citenamefont {Haskell}(1953)}]{Haskell1953}%
  \BibitemOpen
  \bibfield  {author} {\bibinfo {author} {\bibfnamefont {N.~A.}\ \bibnamefont
  {Haskell}},\ }\bibfield  {title} {\bibinfo {title} {{The dispersion of
  surface waves on multilayered media}},\ }\href
  {https://doi.org/10.1785/BSSA0430010017} {\bibfield  {journal} {\bibinfo
  {journal} {Bulletin of the Seismological Society of America}\ }\textbf
  {\bibinfo {volume} {43}},\ \bibinfo {pages} {17} (\bibinfo {year}
  {1953})}\BibitemShut {NoStop}%
\bibitem [{\citenamefont {Potel}\ and\ \citenamefont {{De
  Belleval}}(1993)}]{Potel1993}%
  \BibitemOpen
  \bibfield  {author} {\bibinfo {author} {\bibfnamefont {C.}~\bibnamefont
  {Potel}}\ and\ \bibinfo {author} {\bibfnamefont {J.~F.}\ \bibnamefont {{De
  Belleval}}},\ }\bibfield  {title} {\bibinfo {title} {{Acoustic propagation in
  anisotropic periodically multilayered media: A method to solve numerical
  instabilities}},\ }\href {https://doi.org/10.1063/1.355324} {\bibfield
  {journal} {\bibinfo  {journal} {Journal of Applied Physics}\ }\textbf
  {\bibinfo {volume} {74}},\ \bibinfo {pages} {2208} (\bibinfo {year}
  {1993})}\BibitemShut {NoStop}%
\bibitem [{\citenamefont {Castainqs}\ and\ \citenamefont
  {Hosten}(1994)}]{Castainqs1994}%
  \BibitemOpen
  \bibfield  {author} {\bibinfo {author} {\bibfnamefont {M.}~\bibnamefont
  {Castainqs}}\ and\ \bibinfo {author} {\bibfnamefont {B.}~\bibnamefont
  {Hosten}},\ }\bibfield  {title} {\bibinfo {title} {{Delta operator technique
  to improve the Thomson-Haskell-method stability for propagation in
  multilayered anisotropic absorbing plates}},\ }\href
  {https://doi.org/10.1121/1.408707} {\bibfield  {journal} {\bibinfo  {journal}
  {The Journal of the Acoustical Society of America}\ }\textbf {\bibinfo
  {volume} {95}},\ \bibinfo {pages} {1931} (\bibinfo {year}
  {1994})}\BibitemShut {NoStop}%
\bibitem [{\citenamefont {Knopoff}(1964)}]{Knopoff1964}%
  \BibitemOpen
  \bibfield  {author} {\bibinfo {author} {\bibfnamefont {L.}~\bibnamefont
  {Knopoff}},\ }\bibfield  {title} {\bibinfo {title} {{A matrix method for
  elastic wave problems}},\ }\href {https://doi.org/10.1785/bssa0540010431}
  {\bibfield  {journal} {\bibinfo  {journal} {Bulletin of the Seismological
  Society of America}\ }\textbf {\bibinfo {volume} {54}},\ \bibinfo {pages}
  {431} (\bibinfo {year} {1964})}\BibitemShut {NoStop}%
\bibitem [{\citenamefont {Randall}(1967)}]{Randall1967}%
  \BibitemOpen
  \bibfield  {author} {\bibinfo {author} {\bibfnamefont {M.~J.}\ \bibnamefont
  {Randall}},\ }\bibfield  {title} {\bibinfo {title} {{Fast programs for
  layered half-space problems}},\ }\href
  {https://doi.org/10.1785/bssa0570061299} {\bibfield  {journal} {\bibinfo
  {journal} {Bulletin of the Seismological Society of America}\ }\textbf
  {\bibinfo {volume} {57}},\ \bibinfo {pages} {1299} (\bibinfo {year}
  {1967})}\BibitemShut {NoStop}%
\bibitem [{\citenamefont {Schmidt}\ and\ \citenamefont
  {Jensen}(1985{\natexlab{a}})}]{Schmidt1985a}%
  \BibitemOpen
  \bibfield  {author} {\bibinfo {author} {\bibfnamefont {H.}~\bibnamefont
  {Schmidt}}\ and\ \bibinfo {author} {\bibfnamefont {F.~B.}\ \bibnamefont
  {Jensen}},\ }\bibfield  {title} {\bibinfo {title} {{Efficient numerical
  solution technique for wave propagation in horizontally stratified
  environments}},\ }\href {https://doi.org/10.1016/0898-1221(85)90166-X}
  {\bibfield  {journal} {\bibinfo  {journal} {Computers and Mathematics with
  Applications}\ }\textbf {\bibinfo {volume} {11}},\ \bibinfo {pages} {699}
  (\bibinfo {year} {1985}{\natexlab{a}})}\BibitemShut {NoStop}%
\bibitem [{\citenamefont {Schmidt}\ and\ \citenamefont
  {Jensen}(1985{\natexlab{b}})}]{Schmidt1985b}%
  \BibitemOpen
  \bibfield  {author} {\bibinfo {author} {\bibfnamefont {H.}~\bibnamefont
  {Schmidt}}\ and\ \bibinfo {author} {\bibfnamefont {F.~B.}\ \bibnamefont
  {Jensen}},\ }\bibfield  {title} {\bibinfo {title} {{A full wave solution for
  propagation in multilayered viscoelastic media with application to Gaussian
  beam reflection at fluid-solid interfaces}},\ }\href
  {https://doi.org/10.1121/1.392050} {\bibfield  {journal} {\bibinfo  {journal}
  {The Journal of the Acoustical Society of America}\ }\textbf {\bibinfo
  {volume} {77}},\ \bibinfo {pages} {813} (\bibinfo {year}
  {1985}{\natexlab{b}})}\BibitemShut {NoStop}%
\bibitem [{\citenamefont {Rokhlin}\ and\ \citenamefont
  {Wang}(2002)}]{Rokhlin2002}%
  \BibitemOpen
  \bibfield  {author} {\bibinfo {author} {\bibfnamefont {S.~I.}\ \bibnamefont
  {Rokhlin}}\ and\ \bibinfo {author} {\bibfnamefont {L.}~\bibnamefont {Wang}},\
  }\bibfield  {title} {\bibinfo {title} {{Stable recursive algorithm for
  elastic wave propagation in layered anisotropic media: Stiffness matrix
  method}},\ }\href {https://doi.org/10.1121/1.1497365} {\bibfield  {journal}
  {\bibinfo  {journal} {The Journal of the Acoustical Society of America}\
  }\textbf {\bibinfo {volume} {112}},\ \bibinfo {pages} {822} (\bibinfo {year}
  {2002})}\BibitemShut {NoStop}%
\bibitem [{\citenamefont {Kausel}\ and\ \citenamefont
  {Ro{\"{e}}sset}(1981)}]{Kausel1981}%
  \BibitemOpen
  \bibfield  {author} {\bibinfo {author} {\bibfnamefont {E.}~\bibnamefont
  {Kausel}}\ and\ \bibinfo {author} {\bibfnamefont {J.~M.}\ \bibnamefont
  {Ro{\"{e}}sset}},\ }\bibfield  {title} {\bibinfo {title} {{Stiffness matrices
  for layered soils}},\ }\href {https://doi.org/10.1785/BSSA0710061743}
  {\bibfield  {journal} {\bibinfo  {journal} {Bulletin of the Seismological
  Society of America}\ }\textbf {\bibinfo {volume} {71}},\ \bibinfo {pages}
  {1743} (\bibinfo {year} {1981})}\BibitemShut {NoStop}%
\bibitem [{\citenamefont {Kennett}(2009)}]{Kennett2009}%
  \BibitemOpen
  \bibfield  {author} {\bibinfo {author} {\bibfnamefont {B.}~\bibnamefont
  {Kennett}},\ }\href {https://doi.org/10.22459/SWPSM.05.2009} {\emph {\bibinfo
  {title} {{Seismic Wave Propagation in Stratified Media}}}},\ \bibinfo
  {edition} {new editio}\ ed.\ (\bibinfo  {publisher} {ANU Press},\ \bibinfo
  {year} {2009})\BibitemShut {NoStop}%
\bibitem [{\citenamefont {Jensen}\ \emph {et~al.}(2011)\citenamefont {Jensen},
  \citenamefont {Kuperman}, \citenamefont {Porter},\ and\ \citenamefont
  {Schmidt}}]{Jensen2011}%
  \BibitemOpen
  \bibfield  {author} {\bibinfo {author} {\bibfnamefont {F.~B.}\ \bibnamefont
  {Jensen}}, \bibinfo {author} {\bibfnamefont {W.~A.}\ \bibnamefont
  {Kuperman}}, \bibinfo {author} {\bibfnamefont {M.~B.}\ \bibnamefont
  {Porter}},\ and\ \bibinfo {author} {\bibfnamefont {H.}~\bibnamefont
  {Schmidt}},\ }\href {https://doi.org/10.1007/978-1-4419-8678-8} {\emph
  {\bibinfo {title} {{Computational Ocean Acoustics}}}},\ \bibinfo {edition}
  {second edi}\ ed.,\ Vol.~\bibinfo {volume} {8}\ (\bibinfo  {publisher}
  {Springer},\ \bibinfo {address} {New York, NY},\ \bibinfo {year}
  {2011})\BibitemShut {NoStop}%
\bibitem [{\citenamefont {Nayfeh}(1995)}]{Nayfeh1995}%
  \BibitemOpen
  \bibfield  {author} {\bibinfo {author} {\bibfnamefont {A.~H.}\ \bibnamefont
  {Nayfeh}},\ }\href@noop {} {\emph {\bibinfo {title} {{Wave propagation in
  layered anisotropic media: With application to composites}}}}\ (\bibinfo
  {publisher} {Elsevier},\ \bibinfo {address} {Amsterdam},\ \bibinfo {year}
  {1995})\BibitemShut {NoStop}%
\bibitem [{\citenamefont {Rokhlin}\ \emph {et~al.}(2011)\citenamefont
  {Rokhlin}, \citenamefont {Chimenti},\ and\ \citenamefont
  {Nagy}}]{rokhlin2011}%
  \BibitemOpen
  \bibfield  {author} {\bibinfo {author} {\bibfnamefont {S.~I.}\ \bibnamefont
  {Rokhlin}}, \bibinfo {author} {\bibfnamefont {D.~E.}\ \bibnamefont
  {Chimenti}},\ and\ \bibinfo {author} {\bibfnamefont {P.~B.}\ \bibnamefont
  {Nagy}},\ }\href {https://doi.org/10.1093/oso/9780195079609.001.0001} {\emph
  {\bibinfo {title} {{Physical ultrasonics of composites}}}}\ (\bibinfo
  {publisher} {Oxford University Press},\ \bibinfo {address} {New York, NY},\
  \bibinfo {year} {2011})\BibitemShut {NoStop}%
\bibitem [{\citenamefont {Lowe}(1995)}]{Lowe1995}%
  \BibitemOpen
  \bibfield  {author} {\bibinfo {author} {\bibfnamefont {M.~J.}\ \bibnamefont
  {Lowe}},\ }\bibfield  {title} {\bibinfo {title} {{Matrix Techniques for
  Modeling Ultrasonic Waves in Multilayered Media}},\ }\href
  {https://doi.org/10.1109/58.393096} {\bibfield  {journal} {\bibinfo
  {journal} {IEEE Transactions on Ultrasonics, Ferroelectrics, and Frequency
  Control}\ }\textbf {\bibinfo {volume} {42}},\ \bibinfo {pages} {525}
  (\bibinfo {year} {1995})}\BibitemShut {NoStop}%
\bibitem [{\citenamefont {Pavlakovic}\ \emph {et~al.}(1997)\citenamefont
  {Pavlakovic}, \citenamefont {Lowe}, \citenamefont {Alleyne},\ and\
  \citenamefont {Cawley}}]{Pavlakovic1997}%
  \BibitemOpen
  \bibfield  {author} {\bibinfo {author} {\bibfnamefont {B.}~\bibnamefont
  {Pavlakovic}}, \bibinfo {author} {\bibfnamefont {M.}~\bibnamefont {Lowe}},
  \bibinfo {author} {\bibfnamefont {D.}~\bibnamefont {Alleyne}},\ and\ \bibinfo
  {author} {\bibfnamefont {P.}~\bibnamefont {Cawley}},\ }\bibfield  {title}
  {\bibinfo {title} {{Disperse: A General Purpose Program for Creating
  Dispersion Curves}},\ }in\ \href
  {https://doi.org/10.1007/978-1-4615-5947-4_24} {\emph {\bibinfo {booktitle}
  {Review of Progress in Quantitative Nondestructive Evaluation}}}\ (\bibinfo
  {year} {1997})\ pp.\ \bibinfo {pages} {185--192}\BibitemShut {NoStop}%
\bibitem [{\citenamefont {Kundu}\ \emph {et~al.}(1985)\citenamefont {Kundu},
  \citenamefont {Mal},\ and\ \citenamefont {Weglein}}]{Kundu1985}%
  \BibitemOpen
  \bibfield  {author} {\bibinfo {author} {\bibfnamefont {T.}~\bibnamefont
  {Kundu}}, \bibinfo {author} {\bibfnamefont {A.~K.}\ \bibnamefont {Mal}},\
  and\ \bibinfo {author} {\bibfnamefont {R.~D.}\ \bibnamefont {Weglein}},\
  }\bibfield  {title} {\bibinfo {title} {{Calculation of the acoustic material
  signature of a layered solid}},\ }\href {https://doi.org/10.1121/1.391907}
  {\bibfield  {journal} {\bibinfo  {journal} {The Journal of the Acoustical
  Society of America}\ }\textbf {\bibinfo {volume} {77}},\ \bibinfo {pages}
  {353} (\bibinfo {year} {1985})}\BibitemShut {NoStop}%
\bibitem [{\citenamefont {Cervenka}\ and\ \citenamefont
  {Challande}(1991)}]{Cervenka1991}%
  \BibitemOpen
  \bibfield  {author} {\bibinfo {author} {\bibfnamefont {P.}~\bibnamefont
  {Cervenka}}\ and\ \bibinfo {author} {\bibfnamefont {P.}~\bibnamefont
  {Challande}},\ }\bibfield  {title} {\bibinfo {title} {{A new efficient
  algorithm to compute the exact reflection and transmission factors for plane
  waves in layered absorbing media (liquids and solids)}},\ }\href
  {https://doi.org/10.1121/1.400993} {\bibfield  {journal} {\bibinfo  {journal}
  {The Journal of the Acoustical Society of America}\ }\textbf {\bibinfo
  {volume} {89}},\ \bibinfo {pages} {1579} (\bibinfo {year}
  {1991})}\BibitemShut {NoStop}%
\bibitem [{\citenamefont {Biot}(1956{\natexlab{a}})}]{Biot1956a}%
  \BibitemOpen
  \bibfield  {author} {\bibinfo {author} {\bibfnamefont {M.~A.}\ \bibnamefont
  {Biot}},\ }\bibfield  {title} {\bibinfo {title} {{Theory of Propagation of
  Elastic Waves in a Fluid-Saturated Porous Solid. I. Low-Frequency Range}},\
  }\href {https://doi.org/10.1121/1.1908239} {\bibfield  {journal} {\bibinfo
  {journal} {The Journal of the Acoustical Society of America}\ }\textbf
  {\bibinfo {volume} {28}},\ \bibinfo {pages} {168} (\bibinfo {year}
  {1956}{\natexlab{a}})}\BibitemShut {NoStop}%
\bibitem [{\citenamefont {Biot}(1956{\natexlab{b}})}]{Biot1956b}%
  \BibitemOpen
  \bibfield  {author} {\bibinfo {author} {\bibfnamefont {M.~A.}\ \bibnamefont
  {Biot}},\ }\bibfield  {title} {\bibinfo {title} {{Theory of Propagation of
  Elastic Waves in a Fluid-Saturated Porous Solid. I. Low-Frequency Range}},\
  }\href {https://doi.org/10.1121/1.1908239} {\bibfield  {journal} {\bibinfo
  {journal} {The Journal of the Acoustical Society of America}\ }\textbf
  {\bibinfo {volume} {28}},\ \bibinfo {pages} {168} (\bibinfo {year}
  {1956}{\natexlab{b}})}\BibitemShut {NoStop}%
\bibitem [{\citenamefont {Biot}\ and\ \citenamefont {Willis}(1957)}]{Biot1957}%
  \BibitemOpen
  \bibfield  {author} {\bibinfo {author} {\bibfnamefont {M.~A.}\ \bibnamefont
  {Biot}}\ and\ \bibinfo {author} {\bibfnamefont {D.~G.}\ \bibnamefont
  {Willis}},\ }\bibfield  {title} {\bibinfo {title} {{The Elastic Coefficients
  of the Theory of Consolidation}},\ }\href
  {https://doi.org/10.1007/978-94-017-1112-8_2} {\bibfield  {journal} {\bibinfo
   {journal} {Journal of Applied Mechanics}\ }\textbf {\bibinfo {volume}
  {24}},\ \bibinfo {pages} {594} (\bibinfo {year} {1957})}\BibitemShut
  {NoStop}%
\bibitem [{\citenamefont {Biot}(1962)}]{Biot1962}%
  \BibitemOpen
  \bibfield  {author} {\bibinfo {author} {\bibfnamefont {M.~A.}\ \bibnamefont
  {Biot}},\ }\bibfield  {title} {\bibinfo {title} {{Mechanics of deformation
  and acoustic propagation in porous media}},\ }\href
  {https://doi.org/10.1063/1.1728759} {\bibfield  {journal} {\bibinfo
  {journal} {Journal of Applied Physics}\ }\textbf {\bibinfo {volume} {33}},\
  \bibinfo {pages} {1482} (\bibinfo {year} {1962})}\BibitemShut {NoStop}%
\bibitem [{\citenamefont {Allard}\ \emph {et~al.}(1989)\citenamefont {Allard},
  \citenamefont {Depollier}, \citenamefont {Rebillard}, \citenamefont
  {Lauriks},\ and\ \citenamefont {Cops}}]{Allard1989}%
  \BibitemOpen
  \bibfield  {author} {\bibinfo {author} {\bibfnamefont {J.~F.}\ \bibnamefont
  {Allard}}, \bibinfo {author} {\bibfnamefont {C.}~\bibnamefont {Depollier}},
  \bibinfo {author} {\bibfnamefont {P.}~\bibnamefont {Rebillard}}, \bibinfo
  {author} {\bibfnamefont {W.}~\bibnamefont {Lauriks}},\ and\ \bibinfo {author}
  {\bibfnamefont {A.}~\bibnamefont {Cops}},\ }\bibfield  {title} {\bibinfo
  {title} {{Inhomogeneous Biot waves in layered media}},\ }\href
  {https://doi.org/10.1063/1.344284} {\bibfield  {journal} {\bibinfo  {journal}
  {Journal of Applied Physics}\ }\textbf {\bibinfo {volume} {66}},\ \bibinfo
  {pages} {2278} (\bibinfo {year} {1989})}\BibitemShut {NoStop}%
\bibitem [{\citenamefont {Jocker}\ \emph {et~al.}(2004)\citenamefont {Jocker},
  \citenamefont {Smeulders}, \citenamefont {Drijkoningen}, \citenamefont
  {van~der Lee},\ and\ \citenamefont {Kalfsbeek}}]{Jocker2004}%
  \BibitemOpen
  \bibfield  {author} {\bibinfo {author} {\bibfnamefont {J.}~\bibnamefont
  {Jocker}}, \bibinfo {author} {\bibfnamefont {D.~M.~J.}\ \bibnamefont
  {Smeulders}}, \bibinfo {author} {\bibfnamefont {G.}~\bibnamefont
  {Drijkoningen}}, \bibinfo {author} {\bibfnamefont {C.}~\bibnamefont {van~der
  Lee}},\ and\ \bibinfo {author} {\bibfnamefont {A.}~\bibnamefont
  {Kalfsbeek}},\ }\bibfield  {title} {\bibinfo {title} {{Matrix propagator
  method for layered porous media: Analytical expressions and stability
  criteria}},\ }\href {https://doi.org/10.1190/1.1778249} {\bibfield  {journal}
  {\bibinfo  {journal} {Geophysics}\ }\textbf {\bibinfo {volume} {69}},\
  \bibinfo {pages} {1071} (\bibinfo {year} {2004})}\BibitemShut {NoStop}%
\bibitem [{\citenamefont {Lauriks}\ \emph {et~al.}(1991)\citenamefont
  {Lauriks}, \citenamefont {Allard}, \citenamefont {Depollier},\ and\
  \citenamefont {Cops}}]{Lauriks1991}%
  \BibitemOpen
  \bibfield  {author} {\bibinfo {author} {\bibfnamefont {W.}~\bibnamefont
  {Lauriks}}, \bibinfo {author} {\bibfnamefont {J.~F.}\ \bibnamefont {Allard}},
  \bibinfo {author} {\bibfnamefont {C.}~\bibnamefont {Depollier}},\ and\
  \bibinfo {author} {\bibfnamefont {A.}~\bibnamefont {Cops}},\ }\bibfield
  {title} {\bibinfo {title} {{Inhomogeneous plane waves in layered materials
  including fluid, solid and porous layers}},\ }\href
  {https://doi.org/10.1016/0165-2125(91)90068-Y} {\bibfield  {journal}
  {\bibinfo  {journal} {Wave Motion}\ }\textbf {\bibinfo {volume} {13}},\
  \bibinfo {pages} {329} (\bibinfo {year} {1991})}\BibitemShut {NoStop}%
\bibitem [{\citenamefont {Brouard}\ \emph {et~al.}(1995)\citenamefont
  {Brouard}, \citenamefont {Lafarge},\ and\ \citenamefont
  {Allard}}]{Brouard1995}%
  \BibitemOpen
  \bibfield  {author} {\bibinfo {author} {\bibfnamefont {B.}~\bibnamefont
  {Brouard}}, \bibinfo {author} {\bibfnamefont {D.}~\bibnamefont {Lafarge}},\
  and\ \bibinfo {author} {\bibfnamefont {J.~F.}\ \bibnamefont {Allard}},\
  }\bibfield  {title} {\bibinfo {title} {{A general method of modelling sound
  propagation in layered media}},\ }\href
  {https://doi.org/10.1006/jsvi.1995.0243} {\bibfield  {journal} {\bibinfo
  {journal} {Journal of Sound and Vibration}\ }\textbf {\bibinfo {volume}
  {183}},\ \bibinfo {pages} {129} (\bibinfo {year} {1995})}\BibitemShut
  {NoStop}%
\bibitem [{\citenamefont {Allard}\ and\ \citenamefont
  {Atalla}(2009)}]{Allard2009}%
  \BibitemOpen
  \bibfield  {author} {\bibinfo {author} {\bibfnamefont {J.~F.}\ \bibnamefont
  {Allard}}\ and\ \bibinfo {author} {\bibfnamefont {N.}~\bibnamefont
  {Atalla}},\ }\href@noop {} {\emph {\bibinfo {title} {{Propagation of sound in
  porous media: modeling sound absorbing materials}}}},\ \bibinfo {edition}
  {2nd}\ ed.\ (\bibinfo  {publisher} {Wiley},\ \bibinfo {address} {Chichester,
  UK},\ \bibinfo {year} {2009})\BibitemShut {NoStop}%
\bibitem [{\citenamefont {Pride}\ \emph {et~al.}(2002)\citenamefont {Pride},
  \citenamefont {Tromeur},\ and\ \citenamefont {Berryman}}]{Pride2002}%
  \BibitemOpen
  \bibfield  {author} {\bibinfo {author} {\bibfnamefont {S.~R.}\ \bibnamefont
  {Pride}}, \bibinfo {author} {\bibfnamefont {E.}~\bibnamefont {Tromeur}},\
  and\ \bibinfo {author} {\bibfnamefont {J.~G.}\ \bibnamefont {Berryman}},\
  }\bibfield  {title} {\bibinfo {title} {{Biot slow-wave effects in stratified
  rock}},\ }\href {https://doi.org/10.1190/1.1451799} {\bibfield  {journal}
  {\bibinfo  {journal} {Geophysics}\ }\textbf {\bibinfo {volume} {67}},\
  \bibinfo {pages} {271} (\bibinfo {year} {2002})}\BibitemShut {NoStop}%
\bibitem [{\citenamefont {Jocker}\ and\ \citenamefont
  {Smeulders}(2009)}]{Jocker2009}%
  \BibitemOpen
  \bibfield  {author} {\bibinfo {author} {\bibfnamefont {J.}~\bibnamefont
  {Jocker}}\ and\ \bibinfo {author} {\bibfnamefont {D.~M.~J.}\ \bibnamefont
  {Smeulders}},\ }\bibfield  {title} {\bibinfo {title} {{Ultrasonic
  measurements on poroelastic slabs: Determination of reflection and
  transmission coefficients and processing for Biot input parameters}},\ }\href
  {https://doi.org/10.1016/j.ultras.2008.10.006} {\bibfield  {journal}
  {\bibinfo  {journal} {Ultrasonics}\ }\textbf {\bibinfo {volume} {49}},\
  \bibinfo {pages} {319} (\bibinfo {year} {2009})}\BibitemShut {NoStop}%
\bibitem [{\citenamefont {Nagy}\ \emph {et~al.}(1990)\citenamefont {Nagy},
  \citenamefont {Adler},\ and\ \citenamefont {Bonner}}]{Nagy1990}%
  \BibitemOpen
  \bibfield  {author} {\bibinfo {author} {\bibfnamefont {P.~B.}\ \bibnamefont
  {Nagy}}, \bibinfo {author} {\bibfnamefont {L.}~\bibnamefont {Adler}},\ and\
  \bibinfo {author} {\bibfnamefont {B.~P.}\ \bibnamefont {Bonner}},\ }\bibfield
   {title} {\bibinfo {title} {{Slow wave propagation in air-filled porous
  materials and natural rocks}},\ }\href {https://doi.org/10.1063/1.102872}
  {\bibfield  {journal} {\bibinfo  {journal} {Applied Physics Letters}\
  }\textbf {\bibinfo {volume} {56}},\ \bibinfo {pages} {2504} (\bibinfo {year}
  {1990})}\BibitemShut {NoStop}%
\bibitem [{\citenamefont {Fellah}\ \emph {et~al.}(2007)\citenamefont {Fellah},
  \citenamefont {Fellah}, \citenamefont {Mitri}, \citenamefont {Sebaa},
  \citenamefont {Depollier},\ and\ \citenamefont {Lauriks}}]{Fellah2007}%
  \BibitemOpen
  \bibfield  {author} {\bibinfo {author} {\bibfnamefont {Z.~E.}\ \bibnamefont
  {Fellah}}, \bibinfo {author} {\bibfnamefont {M.}~\bibnamefont {Fellah}},
  \bibinfo {author} {\bibfnamefont {F.~G.}\ \bibnamefont {Mitri}}, \bibinfo
  {author} {\bibfnamefont {N.}~\bibnamefont {Sebaa}}, \bibinfo {author}
  {\bibfnamefont {C.}~\bibnamefont {Depollier}},\ and\ \bibinfo {author}
  {\bibfnamefont {W.}~\bibnamefont {Lauriks}},\ }\bibfield  {title} {\bibinfo
  {title} {{Measuring permeability of porous materials at low frequency range
  via acoustic transmitted waves}},\ }\href {https://doi.org/10.1063/1.2804127}
  {\bibfield  {journal} {\bibinfo  {journal} {Review of Scientific
  Instruments}\ }\textbf {\bibinfo {volume} {78}},\ \bibinfo {pages} {114902}
  (\bibinfo {year} {2007})}\BibitemShut {NoStop}%
\bibitem [{\citenamefont {Fellah}\ \emph {et~al.}(2010)\citenamefont {Fellah},
  \citenamefont {Sebaa}, \citenamefont {Fellah}, \citenamefont {Mitri},
  \citenamefont {Ogam},\ and\ \citenamefont {Depollier}}]{Fellah2010}%
  \BibitemOpen
  \bibfield  {author} {\bibinfo {author} {\bibfnamefont {Z.~E.}\ \bibnamefont
  {Fellah}}, \bibinfo {author} {\bibfnamefont {N.}~\bibnamefont {Sebaa}},
  \bibinfo {author} {\bibfnamefont {M.}~\bibnamefont {Fellah}}, \bibinfo
  {author} {\bibfnamefont {F.~G.}\ \bibnamefont {Mitri}}, \bibinfo {author}
  {\bibfnamefont {E.}~\bibnamefont {Ogam}},\ and\ \bibinfo {author}
  {\bibfnamefont {C.}~\bibnamefont {Depollier}},\ }\bibfield  {title} {\bibinfo
  {title} {{Ultrasonic characterization of air-saturated double-layered porous
  media in time domain}},\ }\href {https://doi.org/10.1063/1.3456443}
  {\bibfield  {journal} {\bibinfo  {journal} {Journal of Applied Physics}\
  }\textbf {\bibinfo {volume} {108}},\ \bibinfo {pages} {014909} (\bibinfo
  {year} {2010})}\BibitemShut {NoStop}%
\bibitem [{\citenamefont {Graff}(1975)}]{Graff1975}%
  \BibitemOpen
  \bibfield  {author} {\bibinfo {author} {\bibfnamefont {K.~F.}\ \bibnamefont
  {Graff}},\ }\href@noop {} {\emph {\bibinfo {title} {{Wave Motion in Elastic
  Solids}}}}\ (\bibinfo  {publisher} {Dover Publications, Inc.},\ \bibinfo
  {address} {New York},\ \bibinfo {year} {1975})\BibitemShut {NoStop}%
\bibitem [{\citenamefont {Morse}\ and\ \citenamefont
  {Freshbach}(1953)}]{Morse1953}%
  \BibitemOpen
  \bibfield  {author} {\bibinfo {author} {\bibfnamefont {P.}~\bibnamefont
  {Morse}}\ and\ \bibinfo {author} {\bibfnamefont {H.}~\bibnamefont
  {Freshbach}},\ }\href@noop {} {\emph {\bibinfo {title} {{Methods of
  Theoretical Physics}}}}\ (\bibinfo  {publisher} {McGraw-Hill Book Company},\
  \bibinfo {year} {1953})\ pp.\ \bibinfo {pages} {52--53}\BibitemShut {NoStop}%
\bibitem [{\citenamefont {Horoshenkov}\ \emph {et~al.}(2016)\citenamefont
  {Horoshenkov}, \citenamefont {Groby},\ and\ \citenamefont
  {Dazel}}]{Horoshenkov2016}%
  \BibitemOpen
  \bibfield  {author} {\bibinfo {author} {\bibfnamefont {K.~V.}\ \bibnamefont
  {Horoshenkov}}, \bibinfo {author} {\bibfnamefont {J.-P.}\ \bibnamefont
  {Groby}},\ and\ \bibinfo {author} {\bibfnamefont {O.}~\bibnamefont {Dazel}},\
  }\bibfield  {title} {\bibinfo {title} {{Asymptotic limits of some models for
  sound propagation in porous media and the assignment of the pore
  characteristic lengths}},\ }\href {https://doi.org/10.1121/1.4947540}
  {\bibfield  {journal} {\bibinfo  {journal} {The Journal of the Acoustical
  Society of America}\ }\textbf {\bibinfo {volume} {139}},\ \bibinfo {pages}
  {2463} (\bibinfo {year} {2016})}\BibitemShut {NoStop}%
\bibitem [{\citenamefont {Horoshenkov}\ \emph {et~al.}(2019)\citenamefont
  {Horoshenkov}, \citenamefont {Hurrell},\ and\ \citenamefont
  {Groby}}]{Horoshenkov2019}%
  \BibitemOpen
  \bibfield  {author} {\bibinfo {author} {\bibfnamefont {K.~V.}\ \bibnamefont
  {Horoshenkov}}, \bibinfo {author} {\bibfnamefont {A.}~\bibnamefont
  {Hurrell}},\ and\ \bibinfo {author} {\bibfnamefont {J.-P.}\ \bibnamefont
  {Groby}},\ }\bibfield  {title} {\bibinfo {title} {{A three-parameter
  analytical model for the acoustical properties of porous media}},\ }\href
  {https://doi.org/10.1121/1.5098778} {\bibfield  {journal} {\bibinfo
  {journal} {The Journal of the Acoustical Society of America}\ }\textbf
  {\bibinfo {volume} {145}},\ \bibinfo {pages} {2512} (\bibinfo {year}
  {2019})}\BibitemShut {NoStop}%
\bibitem [{\citenamefont {Johnson}\ and\ \citenamefont
  {Plona}(1982)}]{Johnson1982a}%
  \BibitemOpen
  \bibfield  {author} {\bibinfo {author} {\bibfnamefont {D.~L.}\ \bibnamefont
  {Johnson}}\ and\ \bibinfo {author} {\bibfnamefont {T.~J.}\ \bibnamefont
  {Plona}},\ }\bibfield  {title} {\bibinfo {title} {{Acoustic slow waves and
  the consolidation transition}},\ }\href {https://doi.org/10.1121/1.388036}
  {\bibfield  {journal} {\bibinfo  {journal} {The Journal of the Acoustical
  Society of America}\ }\textbf {\bibinfo {volume} {72}},\ \bibinfo {pages}
  {556} (\bibinfo {year} {1982})}\BibitemShut {NoStop}%
\bibitem [{\citenamefont {Johnson}\ \emph {et~al.}(1987)\citenamefont
  {Johnson}, \citenamefont {Koplik},\ and\ \citenamefont
  {Dashen}}]{Johnson1987}%
  \BibitemOpen
  \bibfield  {author} {\bibinfo {author} {\bibfnamefont {D.~L.}\ \bibnamefont
  {Johnson}}, \bibinfo {author} {\bibfnamefont {J.}~\bibnamefont {Koplik}},\
  and\ \bibinfo {author} {\bibfnamefont {R.}~\bibnamefont {Dashen}},\
  }\bibfield  {title} {\bibinfo {title} {{Theory of dynamic permeability and
  tortuosity in fluid saturated porous media}},\ }\href
  {https://doi.org/10.1017/S0022112087000727} {\bibfield  {journal} {\bibinfo
  {journal} {Journal of Fluid Mechanics}\ }\textbf {\bibinfo {volume} {176}},\
  \bibinfo {pages} {379} (\bibinfo {year} {1987})}\BibitemShut {NoStop}%
\bibitem [{\citenamefont {Almeida}\ \emph {et~al.}(2023)\citenamefont
  {Almeida}, \citenamefont {Huang}, \citenamefont {Huang}, \citenamefont
  {Cegla},\ and\ \citenamefont {Lan}}]{Almeida2023}%
  \BibitemOpen
  \bibfield  {author} {\bibinfo {author} {\bibfnamefont {A.}~\bibnamefont
  {Almeida}}, \bibinfo {author} {\bibfnamefont {M.}~\bibnamefont {Huang}},
  \bibinfo {author} {\bibfnamefont {P.}~\bibnamefont {Huang}}, \bibinfo
  {author} {\bibfnamefont {F.}~\bibnamefont {Cegla}},\ and\ \bibinfo {author}
  {\bibfnamefont {B.}~\bibnamefont {Lan}},\ }\bibfield  {title} {\bibinfo
  {title} {{Frustrated total internal reflection of ultrasonic waves at a
  fluid-coupled elastic plate}},\ }\bibfield  {journal} {\bibinfo  {journal}
  {arXiv}\ }\href {https://doi.org/10.48550/arXiv.2302.08826}
  {10.48550/arXiv.2302.08826} (\bibinfo {year} {2023})\BibitemShut {NoStop}%
\bibitem [{\citenamefont {Johnson}\ \emph {et~al.}(1982)\citenamefont
  {Johnson}, \citenamefont {Plona}, \citenamefont {Scala}, \citenamefont
  {Pasierb},\ and\ \citenamefont {Kojima}}]{Johnson1982}%
  \BibitemOpen
  \bibfield  {author} {\bibinfo {author} {\bibfnamefont {D.~L.}\ \bibnamefont
  {Johnson}}, \bibinfo {author} {\bibfnamefont {T.~J.}\ \bibnamefont {Plona}},
  \bibinfo {author} {\bibfnamefont {C.}~\bibnamefont {Scala}}, \bibinfo
  {author} {\bibfnamefont {F.}~\bibnamefont {Pasierb}},\ and\ \bibinfo {author}
  {\bibfnamefont {H.}~\bibnamefont {Kojima}},\ }\bibfield  {title} {\bibinfo
  {title} {{Tortuosity and acoustic slow waves}},\ }\href
  {https://doi.org/10.1103/PhysRevLett.49.1840} {\bibfield  {journal} {\bibinfo
   {journal} {Physical Review Letters}\ }\textbf {\bibinfo {volume} {49}},\
  \bibinfo {pages} {1840} (\bibinfo {year} {1982})}\BibitemShut {NoStop}%
\bibitem [{\citenamefont {Liu}\ \emph {et~al.}(1990)\citenamefont {Liu},
  \citenamefont {Ye}, \citenamefont {Weitz},\ and\ \citenamefont
  {Sheng}}]{Liu1990}%
  \BibitemOpen
  \bibfield  {author} {\bibinfo {author} {\bibfnamefont {J.}~\bibnamefont
  {Liu}}, \bibinfo {author} {\bibfnamefont {L.}~\bibnamefont {Ye}}, \bibinfo
  {author} {\bibfnamefont {D.~A.}\ \bibnamefont {Weitz}},\ and\ \bibinfo
  {author} {\bibfnamefont {P.}~\bibnamefont {Sheng}},\ }\bibfield  {title}
  {\bibinfo {title} {{Novel acoustic excitations in suspensions of hard-sphere
  colloids}},\ }\href {https://doi.org/10.1103/PhysRevLett.65.2602} {\bibfield
  {journal} {\bibinfo  {journal} {Physical Review Letters}\ }\textbf {\bibinfo
  {volume} {65}},\ \bibinfo {pages} {2602} (\bibinfo {year}
  {1990})}\BibitemShut {NoStop}%
\bibitem [{\citenamefont {Hu}\ \emph {et~al.}(2008)\citenamefont {Hu},
  \citenamefont {Strybulevych}, \citenamefont {Page}, \citenamefont
  {Skipetrov},\ and\ \citenamefont {van Tiggelen}}]{Hu2008}%
  \BibitemOpen
  \bibfield  {author} {\bibinfo {author} {\bibfnamefont {H.}~\bibnamefont
  {Hu}}, \bibinfo {author} {\bibfnamefont {A.}~\bibnamefont {Strybulevych}},
  \bibinfo {author} {\bibfnamefont {J.~H.}\ \bibnamefont {Page}}, \bibinfo
  {author} {\bibfnamefont {S.~E.}\ \bibnamefont {Skipetrov}},\ and\ \bibinfo
  {author} {\bibfnamefont {B.~A.}\ \bibnamefont {van Tiggelen}},\ }\bibfield
  {title} {\bibinfo {title} {{Localization of ultrasound in a three-dimensional
  elastic network}},\ }\href {https://doi.org/10.1038/nphys1101} {\bibfield
  {journal} {\bibinfo  {journal} {Nature Physics}\ }\textbf {\bibinfo {volume}
  {4}},\ \bibinfo {pages} {945} (\bibinfo {year} {2008})}\BibitemShut {NoStop}%
\bibitem [{\citenamefont {Ecker}\ \emph {et~al.}(2015)\citenamefont {Ecker},
  \citenamefont {Tran}, \citenamefont {Dechent}, \citenamefont {K{\"{a}}bitz},
  \citenamefont {Warnecke},\ and\ \citenamefont {Sauer}}]{Ecker2015}%
  \BibitemOpen
  \bibfield  {author} {\bibinfo {author} {\bibfnamefont {M.}~\bibnamefont
  {Ecker}}, \bibinfo {author} {\bibfnamefont {T.~K.~D.}\ \bibnamefont {Tran}},
  \bibinfo {author} {\bibfnamefont {P.}~\bibnamefont {Dechent}}, \bibinfo
  {author} {\bibfnamefont {S.}~\bibnamefont {K{\"{a}}bitz}}, \bibinfo {author}
  {\bibfnamefont {A.}~\bibnamefont {Warnecke}},\ and\ \bibinfo {author}
  {\bibfnamefont {D.~U.}\ \bibnamefont {Sauer}},\ }\bibfield  {title} {\bibinfo
  {title} {{Parameterization of a Physico-Chemical Model of a Lithium-Ion
  Battery: I. Determination of Parameters}},\ }\href
  {https://doi.org/10.1149/2.0551509jes} {\bibfield  {journal} {\bibinfo
  {journal} {Journal of The Electrochemical Society}\ }\textbf {\bibinfo
  {volume} {162}},\ \bibinfo {pages} {A1836} (\bibinfo {year}
  {2015})}\BibitemShut {NoStop}%
\bibitem [{\citenamefont {Davies}\ \emph {et~al.}(2017)\citenamefont {Davies},
  \citenamefont {Knehr}, \citenamefont {{Van Tassell}}, \citenamefont {Hodson},
  \citenamefont {Biswas}, \citenamefont {Hsieh},\ and\ \citenamefont
  {Steingart}}]{Davies2017}%
  \BibitemOpen
  \bibfield  {author} {\bibinfo {author} {\bibfnamefont {G.}~\bibnamefont
  {Davies}}, \bibinfo {author} {\bibfnamefont {K.~W.}\ \bibnamefont {Knehr}},
  \bibinfo {author} {\bibfnamefont {B.}~\bibnamefont {{Van Tassell}}}, \bibinfo
  {author} {\bibfnamefont {T.}~\bibnamefont {Hodson}}, \bibinfo {author}
  {\bibfnamefont {S.}~\bibnamefont {Biswas}}, \bibinfo {author} {\bibfnamefont
  {A.~G.}\ \bibnamefont {Hsieh}},\ and\ \bibinfo {author} {\bibfnamefont
  {D.~A.}\ \bibnamefont {Steingart}},\ }\bibfield  {title} {\bibinfo {title}
  {{State of Charge and State of Health Estimation Using Electrochemical
  Acoustic Time of Flight Analysis}},\ }\href
  {https://doi.org/10.1149/2.1411712jes} {\bibfield  {journal} {\bibinfo
  {journal} {Journal of The Electrochemical Society}\ }\textbf {\bibinfo
  {volume} {164}},\ \bibinfo {pages} {A2746} (\bibinfo {year}
  {2017})}\BibitemShut {NoStop}%
\bibitem [{\citenamefont {Gold}\ \emph {et~al.}(2017)\citenamefont {Gold},
  \citenamefont {Bach}, \citenamefont {Virsik}, \citenamefont {Schmitt},
  \citenamefont {M{\"{u}}ller}, \citenamefont {Staab},\ and\ \citenamefont
  {Sextl}}]{Gold2017}%
  \BibitemOpen
  \bibfield  {author} {\bibinfo {author} {\bibfnamefont {L.}~\bibnamefont
  {Gold}}, \bibinfo {author} {\bibfnamefont {T.}~\bibnamefont {Bach}}, \bibinfo
  {author} {\bibfnamefont {W.}~\bibnamefont {Virsik}}, \bibinfo {author}
  {\bibfnamefont {A.}~\bibnamefont {Schmitt}}, \bibinfo {author} {\bibfnamefont
  {J.}~\bibnamefont {M{\"{u}}ller}}, \bibinfo {author} {\bibfnamefont {T.~E.}\
  \bibnamefont {Staab}},\ and\ \bibinfo {author} {\bibfnamefont
  {G.}~\bibnamefont {Sextl}},\ }\bibfield  {title} {\bibinfo {title} {{Probing
  lithium-ion batteries' state-of-charge using ultrasonic transmission –
  Concept and laboratory testing}},\ }\href
  {https://doi.org/10.1016/j.jpowsour.2017.01.090} {\bibfield  {journal}
  {\bibinfo  {journal} {Journal of Power Sources}\ }\textbf {\bibinfo {volume}
  {343}},\ \bibinfo {pages} {536} (\bibinfo {year} {2017})}\BibitemShut
  {NoStop}%
\end{thebibliography}%

\end{document}